\documentclass[12pt]{article}
\usepackage{a4wide} 

\RequirePackage[latin1]{inputenc}    

\usepackage[english]{babel}
\usepackage{amsmath}
\usepackage{amsfonts}
\usepackage{amssymb}
\usepackage{xspace}
\usepackage{latexsym}
\usepackage{url}
\usepackage{xspace}
\usepackage{fancyvrb}

\usepackage{graphics,color}             
\RequirePackage[latin1]{inputenc}  

\newenvironment{restate-proposition}[2][{}]{\noindent\textbf{Proposition~{#2}}\;\textbf{#1}\ 
}{\vskip 1em}

\newenvironment{restate-theorem}[2][{}]{\noindent\textbf{Theorem~{#2}}\;\textbf{#1}\ 
}{\vskip 1em}

\newenvironment{restate-corollary}[2][{}]{\noindent\textbf{Corollary~{#2}}\;\textbf{#1}\ 
}{\vskip 1em}

\newcommand{\Proofitemb}[1]{\medskip \noindent {\bf #1\;}}
\newcommand{\Proofitemfb}[1]{\noindent {\bf #1\;}}
\newcommand{\Proofitem}[1]{\medskip \noindent $#1\;$}
\newcommand{\Proofitemf}[1]{\noindent $#1\;$}
\newcommand{\Defitem}[1]{\smallskip \noindent $#1\;$}
\newcommand{\Defitemf}[1]{\noindent $#1\;$}

\makeatother

\makeatletter
\def\@ysproof[#1]{\@sproof{ #1}}
\def\@sproof#1{\begin{trivlist}\item[]{\textit{Sketch of the proof#1.}}}

\makeatletter
\def\@yproof[#1]{\@proof{ #1}}
\def\@proof#1{\begin{trivlist}\item[]{\textit{Proof#1.}}}

\newcommand{\hbra}{\noindent\hbox to \textwidth{\leaders\hrule height1.8mm depth-1.5mm\hfill}}
\newcommand{\hket}{\noindent\hbox to \textwidth{\leaders\hrule height0.3mm\hfill}}
\newcommand{\ratio}{.3}

\newtheorem{theorem}{Theorem}

\newtheorem{definition}[theorem]{Definition}
\newtheorem{lemma}[theorem]{Lemma}

\newtheorem{proposition}[theorem]{Proposition}
\newtheorem{example}[theorem]{Example}

\newtheorem{remark}[theorem]{Remark}

\newcommand{\Proof}{\noindent {\sc Proof}. }

\newcommand{\qed}{\hfill${\Box}$}

\newcommand{\Figbar}{{\center \rule{\hsize}{0.3mm}}}    

\newcommand{\cl}[1]{{\cal #1}}          

\newcommand{\lf}{\lfloor}
\newcommand{\rf}{\rfloor}
\newcommand{\ol}[1]{\overline{#1}}      

\newcommand{\arrow}{\rightarrow}        
\newcommand{\trarrow}{\stackrel{*}{\rightarrow}}        
\newcommand{\Nat}{\mathbf{N}}                 
\newcommand{\Alt}{ \mid\!\!\mid  }

\newcommand{\dpar}{\mid\!\mid}
\newcommand{\infer}[2]{\begin{array}{c} #1 \\ \hline #2 \end{array}}

\newcommand{\Arrow}{\Rightarrow}        

\newcommand{\dl}{[\![}                  
\newcommand{\dr}{]\!]}                  

\newcommand{\union}{\cup}               
\newcommand{\inter}{\cap}               
\newcommand{\minus}{\backslash}         
\newcommand{\comp}{\circ}               
\newcommand{\set}[1]{\{#1\}}            
\newcommand{\pset}[1]{\{\! | #1 |\!\}}  

\newcommand{\join}{\vee}                
\newcommand{\dcl}{\downarrow}           
\newcommand{\ucl}{\uparrow}             
\newcommand{\rel}[1]{\;{\cal #1}\;}     
\newcommand{\rl}[1]{\;#1\;}             

\newcommand{\true}{{\it true}}                             
\newcommand{\false}{{\it false}}                            

\newcommand{\mand}{\mbox{ and }}
\newcommand{\w}[1]{{\it #1}}    

\newcommand{\lc}{\langle\!|}
\newcommand{\rc}{|\!\rangle}

\newcommand{\qqs}[2]{\forall\, #1\;\: #2}
\newcommand{\xst}[2]{\exists\, #1\;\: #2}

\newcommand{\s}[1]{{\sf #1}}    
\newcommand{\vc}[1]{{\bf #1}}

\newcommand{\act}[1]{\stackrel{#1}{\rightarrow}} 
\newcommand{\wact}[1]{\stackrel{#1}{\Rightarrow}} 

\newcommand{\sbis}{\equiv_L}

\newcommand{\susp}{\downarrow}
\newcommand{\lsusp}{\Downarrow_L}
\newcommand{\wsusp}{\Downarrow}

\newcommand{\bbis}{\approx_{B}}
\newcommand{\cbis}{\approx_{C}}
\newcommand{\lbis}{\approx_{L}}

\newcommand{\lbissusp}{\approx_{L}^{\susp}} 

\begin{document}

\title{The SL synchronous language, revisited}

\author{Roberto M. Amadio\thanks{Partially supported by ACI {\em S\'ecurit\'e Informatique} CRISS.}\\
        Universit\'e Paris 7\thanks{Laboratoire {\em Preuves,
        Programmes et Syst\`emes}, UMR-CNRS 7126.}}

\maketitle

\begin{abstract}
We revisit the SL synchronous programming model introduced by
Boussinot and De Simone {\em (IEEE, Trans. on Soft. Eng., 1996)}. 
We discuss an alternative design of the model including {\em thread
spawning} and {\em recursive definitions} and
we explore some basic properties of the revised model:
determinism, reactivity, CPS translation to a tail recursive form,
computational expressivity, and a compositional notion of
program equivalence.
\end{abstract}

\section{Introduction}
In synchronous models the computation of a set of participants is
regulated by a notion of {\em instant}.  The {\em Synchronous
Language} introduced in \cite{BD95} belongs to this category.  A {\em
program} in this language generally contains sub-programs running in
parallel and interacting via shared {\em signals}.  By default, at the
beginning of each instant a signal is absent and once it is emitted it
remains in that state till the end of the instant.  The model can be
regarded as a relaxation of the {\sc Esterel} model \cite{BG92} where
the {\em reaction to the absence} of a signal is delayed to the
following instant, thus avoiding the difficult problems due to {\em
causality cycles} in {\sc Esterel} programs.

The model has gradually evolved into a programming language for
concurrent applications and has been implemented in the context of
various programming languages such as \textsc{C}, \textsc{Java},
\textsc{Scheme}, and \textsc{Caml} (see, {\em e.g.},
\cite{mimosarp,SchemeFT,MandelPouzetPPDP05}).  The design accommodates
a dynamic computing environment with threads entering or leaving the
synchronisation space \cite{Boudol04}. In this context, it seems
natural to suppose that the scheduling of the threads is only
determined at run time (as opposed to certain synchronous languages
such as {\sc Esterel} or {\sc Lustre}).  It appears that many typical
``concurrent'' applications such as event-driven controllers, data
flow architectures, graphical user interfaces, simulations, web
services, multiplayer games, are more effectively programmed in the
synchronous framework.

The SL language was carefully designed to be compiled to finite state
automata. Motivated by the evolution of the language mentioned above,
we consider a {\em synchronous language} including {\em thread spawning},
and {\em recursive definitions} (section \ref{model-sec}) and
we explore some basic properties of the revised model.
First, we prove that the resulting language is deterministic and 
provide a simple static analysis that entails reactivity (section \ref{basic-properties-sec}).
Second, we propose a continuation passing style translation to
a more basic language of tail recursive threads (section \ref{cps-sec}).
Third, we show that the language without signal generation
has the same computational power as a class of `monotonic' Mealy
machines, while the language with signal generation is Turing
equivalent (section \ref{expressivity-sec}).
Fourth, we introduce a notion of contextual barbed bisimulation
and characterise it via a suitable labelled bisimulation (section
\ref{equiv-sec}). Some standard proofs are delayed to the appendix
\ref{proof-appendix}.

\subsection{Related work}
This work is a continuation of \cite{ABBC05} where we 
outline results and problems connected with the SL model
10 years after its proposal.
A determinacy theorem was already stated in the original paper
\cite{BD95} with a similar proof based on the confluence of the
`small step' reduction. Of course, many other determinacy theorems
occur in the literature on synchronous programming (cf., {\em e.g.},
\cite{Kahn74}).
The static analysis technique for ensuring reactivity is
inspired by previous work by the author \cite{AD04,AD05}
where, roughly, the reactivity of a (tail recursive) SL model 
with data types is studied.
The tail recursive SL model and the related CPS translation
appear to be original. They arose out of an attempt 
to understand the relative expressivity
of various synchronous operators such as $\s{await}$,
$\s{when}$ and $\s{watch}$.
The results on the computational expressivity of the revised model,
notably its characterisation via monotonic Mealy machine,
were motivated by the compilation to finite
state machines in the original SL proposal \cite{BD95}.
Finally, there seems to be no previous attempt at developing
a compositional notion of bisimulation equivalence for the
SL model in a CCS style. However a specific notion of 
bisimulation for `closed systems' has been proposed recently in the
framework of the work on non-interference
for synchronous systems \cite{BCM04}.


\section{The model}\label{model-sec}
In this section, we present a formalisation of the model which is
largely inspired by the original proposition \cite{BD95} and a recent
survey \cite{ABBC05}. We anticipate that in section \ref{cps-sec} we
will simplify the control structure by moving to a tail recursive
model and in section \ref{equiv-sec} we will discuss an alternative
presentation in the spirit of process calculi.
%

\subsection{Environments}
We assume a countable set $S$ of {\em signal names} $s,s',\ldots$.  We
suppose a subset $\w{Int}=\w{Input}\union \w{Output}$ of $S$ of {\em
 observable} signal names representing input or output signals and such that $S\minus
\w{Int}$ is infinite.
An {\em environment} $E$ is a partial function from
signal names to boolean values $\true$ and $\false$ whose domain
of definition $\w{dom}(E)$  contains $\w{Int}$ and such that 
$S\minus \w{dom}(E)$ is infinite.  

\subsection{Threads}\label{threads-sec}
We denote with $\vc{x}$ a vector of elements $x_1,\ldots,x_n$, $n\geq
0$ and with $[\_/\_]$ the usual substitution. By default, 
bound names can be renamed.
We denote with $A(\vc{s}), B(\vc{s}),\ldots$ thread identifiers with
parameters $\vc{s}$. As usual, each thread identifier is defined by exactly one
equation $A(\vc{x})=T$ where $T$ is a {\em thread} defined by the grammar:
\[
T::=0 \Alt (T;T) \Alt (\s{emit}\ s) \Alt  
           (\nu s\ T) \Alt (\s{thread}\ T) \Alt
           (\s{await} \ s) \Alt (\s{watch} \ s \ T) \Alt A(\vc{s})
\]
and the signal names free in $T$ are contained in $\set{\vc{x}}$.
Sometimes, some of the parameters (possibly all) are fixed and 
in these cases we will feel free to omit them.
A thread is executed relatively to an environment which is {\em
shared} with other parallel threads.  The intended semantics is as
follows: $0$ is the terminated thread; 
 $T;T$ is the usual sequentialisation;
$(\s{emit}\ s)$ emits $s$, {\em
i.e.}  sets to $\true$ the signal $s$ and terminates, $(\nu 
s\ T)$ creates a fresh signal which is local to the thread $T$ ($s$ is
bound in $T$) and executes $T$; $(\s{thread}\ T)$ 
spawns a thread $T$ which will be
executed in parallel and terminates; $(\s{await} \ s)$ terminates
if the signal $s$ is present and suspends the
execution otherwise; $(\s{watch} \ s \ T)$ allows the execution of $T$
but terminates $T$ at the end of the first instant where the signal
$s$ is present. The implementation of the \s{watch} instruction 
requires to stack the signals that may cause the abortion of the
current thread together with the associated continuations.
For instance, in 
$(\s{watch} \ s_1 \ (\s{watch} \ s_2 \ T_1);T_2);T_3$,
we start executing $T_1$. Assuming
that at the end of the instant, the execution of $T_1$ is not
completed, the computation in the following instant resumes with $T_3$
if $s_1$ was present at the end of the instant, with $T_2$ if $s_1$
was absent and $s_2$ was present at the end of the instant, and with
the residual of $T_1$, otherwise.  
We point out that a thread spawned by the \s{thread} instruction,
escapes the \s{watch} signals and the related continuations.

\subsection{Thread reduction}
A {\em program} $P$ is a finite non-empty {\em multi-set} of threads.  
We denote with $\w{sig}(T)$ ($\w{sig}(P)$) the set of signals free in $T$
(in threads in $P$).  Whenever we write $(T,E)$, $(P,E)$ it is
intended that $\w{sig}(T)\subseteq \w{dom}(E)$, $\w{sig}(P)\subseteq
\w{dom}(E)$, respectively.  All reduction 
rules maintain the invariant that the signals defined in the thread or in
the program are in the domain of definition of the associated
environment.  In particular, all signal names which are not in the
domain of definition of the environment are guaranteed to be {\em
fresh}, i.e., not used elsewhere in the program.
Finally, we make the usual assumption that reduction 
rules are given modulo renaming of the bound signal names.

We assume that sequential composition `;' associates to the right.
A {\em redex} $\Delta$ is defined by the grammar:
\[
\Delta::= 0;T \Alt (\s{emit} \ s) \Alt 
          (\nu s\ T) \Alt (\s{thread}\ T) \Alt 
          (\s{await} \ s) \Alt (\s{watch} \ s \ 0) \Alt A(\vc{s})~.
\]
An {\em evaluation context} $C$ is defined by the grammar:
\[
C::= [\ ] \Alt [ \ ];T \Alt (\s{watch} \ s \ C) \Alt (\s{watch} \ s \ C);T~.
\]
We have a canonical decomposition of a thread in an evaluation context
and a redex whose proof is delayed to appendix \ref{proof-prop-unique-decomp}.

\begin{proposition}[unique decomposition]\label{prop-unique-decomp}
A thread $T\neq 0$ admits a unique decomposition
$T=C[\Delta]$ into an evaluation context $C$ and a redex $\Delta$.
Moreover, if $T=0$ then no decomposition exists.
\end{proposition}

The reduction relation $(T,E)\act{P} (T',E')$ 
is defined first on redexes by the rules
$(T_{1-7})$ and then it is lifted to threads 
by the rule $(T_8)$:
\[
\begin{array}{llll}
(T_1) &(0;T,E) &\act{\emptyset} (T,E) \\

(T_2) &(\s{emit} \ s,E) &\act{\emptyset} (0,E[\true/s]) \\

(T_3) &(\s{watch} \ s \ 0,E) &\act{\emptyset} (0,E) \\

(T_4) &(\nu  s \ T,E) &\act{\emptyset} (T,E[\false/s]) 
      &\mbox{if  }s\notin \w{dom}(E) \\

(T_5) &(A(\vc{s}),E) &\act{\emptyset} ([\vc{s}/\vc{x}]T,E) 
      &\mbox{if  }A(\vc{x})=T \\ 

(T_6) &(\s{await} \ s ,E)
      &\act{\emptyset} (0,E) 
      &\mbox{if  }E(s)=\true \\ 

(T_7) &(\s{thread} \ T,E) 
      &\act{\pset{T}} (0,E)  \\

(T_8) &(C[\Delta],E) &\act{P} (C[T'],E')
      &\mbox{if }(\Delta,E) \act{P} (T',E')
\end{array}
\]
We write $(T,E)\susp$ if $T$ cannot be reduced in 
the environment $E$ according to the rules above. 
We also say that $(T,E)$ is {\em suspended}.
An inspection of the rules reveals that 
$(T,E)\susp$ if and only if
$T=0$ or $T=C[(\s{await} \ s)]$ with $E(s)=\false$.
Thus the $\s{await}$ statement is the only one that
may cause the suspension of a thread.
The suspension predicate is extended to programs as follows
$(P,E)\susp$ if $\qqs{T\in P}{(T,E)\susp}$.

\subsection{Program reduction}\label{program-sec}
To execute a program $P$ in an environment $E$ during an instant
proceed as follows:  

\Defitem{(1)} Schedule 
(non-deterministically) the executions of the threads
that compose it as long as some progress is possible according
to the rule:
\[
(P\union \pset{T},E) \arrow (P\union \pset{T'} \union P'',E')
\quad \mbox{ if } \quad (T,E)\act{P''} (T',E') ~.
\]
We also write $(P\union \pset{T},E)\act{P''} (P\union \pset{T'},E')$
if $(T,E)\act{P''} (T',E')$.

\Defitem{(2)}
Transform all $(\s{watch}\ s\ T)$ instructions where
the signal $s$ is present into the terminated thread $0$.
Formally, we
rely on the function $\lf \_ \rf_E$ defined on a multiset of 
suspended threads as follows:
\[
\begin{array}{c}
\lf P \rf_E = \pset{\lf T \rf_E \mid T\in P} 

\quad 
\lf 0 \rf_E = 0 

\quad
\lf T;T' \rf_E = \lf T \rf_E ; T'  

\quad 
\lf \s{await} \ s \rf_E = (\s{await} \ s)

\\[+.5em]

\lf \s{watch}\ s \ T \rf_E = 
\left\{ 
\begin{array}{ll}
0 &\mbox{if } E(s)=\true \\
(\s{watch}\ s \ \lf T \rf_E) &\mbox{otherwise}
\end{array} \right.

\end{array}
\]

\subsection{Trace semantics}\label{trace-semantics}
Finally, the input-output behaviour of a program is 
described by labelled transitions
$P\act{I/O} P'$ where $I\subseteq \w{Input}$ and
$O\subseteq \w{Output}$ are the signals in the
interface which are present in input at the beginning of the instant and 
in output at the end of the instant, respectively.  
As in Mealy machines, the transition means that from program (state) $P$ with 
`input' signals $I$ we move to program (state) $P'$ with `output'
signals $O$. This is formalised by the rule:
\[
\begin{array}{c}

(I/O)
\quad 
\infer{(P,E_{I,P}) \trarrow (P', E'), \quad  (P',E')\susp,\quad
O=\set{s\in \w{Output} \mid E'(s)=\true}}
      {P\act{I/O} P'}  \\[+.5em]

\mbox{where:}
\quad
E_{I,P}(s)= \left\{ 
      \begin{array}{ll}
      \true  &\mbox{if }s\in I \\
      \false &\mbox{if }s\in(\w{Int}\union \w{sig}(P))\minus I \\
      \w{undefined}      &\mbox{otherwise}
      \end{array} \right. 
\end{array}
\]
Note that in the definition of $E_{I,P}$ 
we insist on having all signals free in the program
in the domain of definition of the environment and we leave the others
undefined so that they can be potentially used in the rule $(T_4)$.
A {\em complete} run of a program $P$ is a reduction
$P\act{I_{1}/O_{1}} P_1 \act{I_{2}/O_{2}} P_2 \cdots$ which
is either infinite or is finite and cannot be further extended.
We define an extensional semantics of a program $P$,
as the set $\w{tr}(P)$ of (finite or infinite) words 
associated with its complete runs. Namely:
\begin{equation}
\w{tr}(P) = \set{ (I_1/O_1)(I_2/O_2)\cdots \mid I_j\subseteq
  \w{Input}, O_j\subseteq \w{Output}, 
   P\act{I_{1}/O_{1}} P_1 \act{I_{2}/O_{2}} P_2 \cdots}
\end{equation}


\subsection{Derived instructions}\label{def-instr}
We may abbreviate $(\nu  s_1 \cdots (\nu  s_n\ T)\cdots)$
as  $(\nu  s_1,\ldots,s_n\ T)$ and
$(\s{thread} \ T_1);\cdots (\s{thread}\ T_n)$ as
$(\s{thread} \ T_1,\ldots,T_n)$.
Table \ref{derived-operators} presents some derived
instructions which are frequently used in the programming
practice.
The instruction $(\s{loop} \ T)$ can be thought as 
$T;T;T;\cdots$. Note that in $(\s{loop} \ T);T'$, $T'$ is
{\em dead code}, {\em i.e.}, it can never be executed.
The instruction $(\s{now} \ T)$ runs $T$ for the current
instant, {\em i.e.}, if the execution of $T$ is not
completed within the current instant then it is aborted.
The instruction $\s{pause}$ suspends the execution of the
thread for the current instant and resumes it in the following
one. We  will rely on this instruction to guarantee 
the termination of the computation of each thread 
within an instant (see section \ref{basic-properties-sec}).
The instruction $(\s{present}\ s \ T_1\ T_2)$ branches
on the presence of a signal. Note that the branch
$T_2$ corresponding to the {\em absence} of the signal 
is executed in the following instant and that
we suppose $s'\notin \w{sig}(T_1)\union \w{sig}(T_2)$. 
The instruction $(T_1 \dpar T_2)$ runs in parallel 
the threads $T_1$ and $T_2$ and waits for their
termination.
Here we suppose that 
$s_1,s_2,s'_{1},s'_{2} \notin \w{sig}(T_1)\union \w{sig}(T_2)$.





\begin{table}
\[
\begin{array}{ll}

(\s{loop} \ T)    &= A\quad\mbox{where:
  }A \ = \ T;A \\

(\s{now} \ T )    &= \nu  s \ (\s{emit} \ s);(\s{watch} \ s \ T)    
                  \quad s\notin\w{sig}(T) \\

\s{pause}         &= \nu  s \ (\s{now} \ (\s{await} \ s)) \\




(\s{present}\ s \ T_1\ T_2)
&= \nu  s'\ (\s{thread} \ \\
&\qquad \qquad  (\s{now} \ (\s{await} \ s);(\s{thread}\ T_1;(\s{emit}\ s'))), \\
&\qquad \qquad  (\s{watch}\ s \ \s{pause};(\s{thread}\ T_2;(\s{emit}\ s'))) \  ); 
(\s{await}\ s') \\

(T_1 \dpar T_2)
&=\nu  s_1, s_2, s'_{1}, s'_{2} \ (\s{thread}  \\
&\qquad\qquad (\s{watch} \ s'_{1} \ T_1; (\s{loop}\ (\s{emit} \ s_1);\s{pause})), \\
&\qquad\qquad (\s{watch} \ s'_{2} \ T_2; (\s{loop}\ (\s{emit}\ s_2);\s{pause}))\ ); \\ 
&\qquad\qquad \quad (\s{await} \ s_1);(\s{emit}\ s'_{1}); (\s{await} \ s_2);(\s{emit} \ s'_{2})


\end{array}
\]
\caption{Some derived instructions}\label{derived-operators}
\end{table}

\subsection{Comparison with \cite{BD95}}
The main novelty with respect to \cite{BD95} is the 
replacement of \s{loop} and parallel composition operators with
recursive definitions and \s{thread} spawning.
We should stress that the encoding of the \s{present}
and parallel composition operators 
do not correspond exactly to the operators in the original language.
This is because the instructions $T_1$ and $T_2$ 
are under a \s{thread} instruction and therefore
their execution does {\em not} depend on watch 
signals that may be on top of them. If this must be the case,
then we must prefix $T_1$ and $T_2$ with suitable 
\s{watch} instructions. The CPS translation discussed
in section \ref{cps-sec}, provides a systematic method to simulate
the stack of watch signals.

\subsection{Cooperative vs. preemptive concurrency}
In {\em cooperative} concurrency a running thread cannot be
interrupted unless it explicitly decides to return the control to the
scheduler. This is to be contrasted with {\em preemptive} concurrency
where a running thread can be interrupted at any point unless it
explicitly requires that a series of actions is atomic ({\em e.g.},
via a lock).  We refer to, {\em e.g.},~\cite{Ous96} for an extended 
comparison of the cooperative and preemptive models in the practice 
of programming.  In its original proposal, the SL language adopts
a cooperative notion of concurrency. Technically this means that
a `big step' reduction is defined on top of the `small step'
reduction we have introduced. The big step reduction runs a thread
atomically till it terminates or it suspends on an \s{await} 
statement. Programs are then evaluated according to this big step
reduction. In particular, this means that the small step 
reductions cannot be freely interleaved. In the following, we will focus on the 
small step/preemptive semantics and neglect the big step/cooperative
semantics for two reasons: (1) All main results (determinism,
reactivity, CPS translation) are naturally obtained at the level of
the small step/preemptive semantics and are then lifted
to the big step/cooperative semantics.
(2) The cooperative semantics goes against 
the natural idea of executing a program with parallel threads 
on a multi-processor where the threads run in parallel on
different processors up to a synchronisation point.

\section{Determinism and reactivity}\label{basic-properties-sec}
We consider two important properties a SL program should have:
{\em determinism}  and {\em reactivity}. While the first
property is ensured by the design of the language (as was the case in the original
language), we enforce the second by means of a new static analysis.

\subsection{Determinism}
It is immediate to verify that the evaluation of a thread $T$ in an
environment $E$ is deterministic.  Therefore the only potential source
of non-determinism comes from the scheduling of the threads.  The
basic remark is that the emission of a signal can
never block the execution of a statement within an instant. The more
signals are emitted the more the computation of a thread can progress
within an instant. Of course, this {\em monotonicity property}
relies on the fact that a thread
cannot detect the absence of a signal before the end of an instant.

Technically, the property that entails determinism is the
fact that the small step reduction is strongly confluent
up to {\em renaming}.
A renaming $\sigma$ is a bijection $\sigma$ on
signal names which is the identity on the names in the
interface $\w{Int}$.
We introduce a notion of {\em equality up to renaming}:
(i) $T=_\alpha T'$ if there is a renaming $\sigma$ such
that $\sigma T = T'$ and 
(ii) $(T,E)=_\alpha (T',E')$ if there is a renaming $\sigma$ such
that $\sigma T = T'$ and $E=E'\comp \sigma$.
In a similar way, we define  $P=_\alpha P'$  and
$(P,E) =_\alpha (P',E')$.
We rely on equality up to renaming to define a notion of
determinism.

\begin{definition}
The set of {\em deterministic} programs is the
largest set of programs $\cl{D}$ such that 
if $P\in \cl{D}$, $I\subseteq \w{Input}$,
$P\act{I/O_{1}} P_1$, and $P\act{I/O_{2}} P_2$
then $O_1 = O_2$ and $P_1 =_\alpha P_2 \in \cl{D}$.
\end{definition}

In appendix \ref{proof-deterministic-prop}, 
we show how to derive determinism from strong confluence 
by means of a standard tiling argument.

\begin{theorem}\label{deterministic-prop}
All programs are deterministic.
\end{theorem}

\subsection{Reactivity}
We now turn to a formal definition of reactivity.

\begin{definition}
The set of {\em reactive} programs is the largest set of programs $\cl{R}$
such that if $P\in \cl{R}$ then for every choice $I\subseteq \w{Input}$ of the
input signals there are $O,P'$ such that $P\act{I/O}P'$ and $P'\in \cl{R}$.
\end{definition}

We can write programs which are not reactive.  For instance, the
thread $A=(\s{await}\ s ); A$ may potentially loop within an instant.
Whenever a thread loops within an instant the computation of the whole
program is blocked as the instant never terminates.  In the
programming practice, reactivity is ensured by instrumenting the code
with \s{pause} statements that force the computation to suspend for
the current instant.
Following this practice, we take the \s{pause}
statement as a primitive, though it can can be defined as seen in section 
\ref{def-instr}. This can be easily done by 
observing that a suspended  thread may also have the shape
$C[\s{pause}]$ and by extending the evaluation at the end of the instant
with the equation $\lf \s{pause} \rf_E = 0$.
We introduce next a {\em static analysis} that guarantees
reactivity on a code with explicit \s{pause} statements.

We denote with $X,Y,\ldots$ finite multisets of 
thread identifiers and with $\ell$ a label ranging
over the symbols $0$ and $\dcl$.
We define a function $\w{Call}$ associating with a thread
$T$ a pair $(X,\ell)$ where intuitively the multi-set $X$ represents
the thread identifiers that $T$ may call within
the current instant and $\ell$ indicates whether a continuation of $T$
has the possibility of running within the current instant $(\ell=0)$
or not $(\ell=\dcl)$. As usual, $\pi_i$ projects a tuple on the
$i^{\w{th}}$ component.
\[
\begin{array}{cc}

\w{Call}(0)=\w{Call}(\s{emit}\ s)= \w{Call}(\s{await} \ s) = (\emptyset,0)

&\w{Call}(\s{pause}) = (\emptyset,\dcl) \\[+.5em]

\w{Call}(\nu  s \ T) = \w{Call}(\s{watch} \ s \ T) = \w{Call}(T)  
&\w{Call}(A(\vc{s})) = (\pset{A},0)  \\[+.5em]

\w{Call}(\s{thread} \ T) = (\pi_1(\w{Call}(T)),0)

&\w{Call}(T_1;T_2) = \w{Call}(T_1);\w{Call}(T_2) 

\end{array}
\]
where the operation `;' is defined on the codomain of $\w{Call}$ as follows:
\[
\begin{array}{l|cc}

;               &(Y,0) &(Y,\dcl) \\\hline

(X,0)           &(X\union Y,0) &(X\union Y,\dcl) \\
(X,\dcl)        &(X,\dcl)      &(X,\dcl) 
\end{array}
\]
We notice that this operation is {\em associative}.
It is convenient to define the \w{Call} function also 
on evaluation contexts as follows:
\[
\begin{array}{cc}

\w{Call}([~]) = \emptyset
&\w{Call}([~];T) = \w{Call}(T) \\
\w{Call}(\s{watch} \ s \ C) = \w{Call}(C)
&\w{Call}((\s{watch} \ s \ C);T') = \w{Call}(C);\w{Call}(T')
\end{array}
\]
and observe the following property which is proved by induction
on the structure of the context.

\begin{proposition}\label{call-cxt}
For every evaluation context $C$ and thread $T$,
$\w{Call}(C[T]) = \w{Call}(T);\w{Call}(C)$.
\end{proposition}

We can now introduce a static condition that guarantees reactivity.
Intuitively, to ensure the reactivity of a program $P$, it is enough to find
an {\em acyclic precedence relation} on the related thread identifiers which
is consistent with their definitions. Namely, we define:
\[
\w{Cnst}(P) = 
\set{A>B \mid A(\vc{x})=T \mbox{ equation for program }P, B\in \pi_1(\w{Call}(T))}
\]

\begin{theorem}\label{reactivity-thm}
A program $P$ is reactive if there is a well founded order $>$ on
thread identifiers that satisfies the inequalities in $\w{Cnst}(P)$.
\end{theorem}
\Proof 
The order $>$ on thread identifiers induces a well founded 
order on the finite multi-sets of thread identifiers.
We denote this order with  $>_{m,{\w{Id}}}$.
We define a {\em size function} \w{sz} from threads to natural number
$\Nat$ as follows:
\[
\begin{array}{c}
\w{sz}(0)=\w{sz}(\s{pause}) = 0, \quad
\w{sz}(\s{emit} \ s) = \w{sz}(\s{await} \ s) = \w{sz}(A(\vc{s})) = 1,
\\
\w{sz}(\nu  s\ T)= \w{sz}(\s{watch} \ s\ T) = 
\w{sz}(\s{thread} \ T) = 1 +\w{sz}(T),

\quad
\w{sz}(T_1;T_2) = 1+ \w{sz}(T_1)+ \w{sz}(T_2)
\end{array}
\]
We denote with $>_{\w{lex}}$ the lexicographic order from left to right 
induced by the order $>_{m,{\w{Id}}}$ and
the standard order on natural numbers. 
This order is well-founded.
Finally, we consider the multi-set order $>_m$ induced by
$>_{\w{lex}}$ on finite multi-sets. Again, this order is well founded.
Next, we define a `measure' $\mu$ associating with a program 
a finite multi-set:
\[
\mu(P) = \pset{(\pi_1(\w{Call}(T)),\w{sz}(T)) \mid T\in P}~.
\]
It just remains to check that the small step reduction
decreases this measure. Namely, 
if $(P,E) \act{P''} (P',E')$ then 
$\mu(P)>_m \mu(P')\union \mu(P'')$, where the $\union$
is of course intended on multi-sets.
We recall that in the multi-set order an element can
be replaced by a finite multi-set of strictly smaller elements.
We proceed by case analysis on the small step reduction.

\Proofitem{\bullet}
Suppose the program reduction is induced by the thread reduction:
\[
(C[\Delta],E)  \act{\emptyset} (C[T],E)~.
\]
where $\Delta$ has the shape 
$0;T'$, $\s{emit} \ s$, $\nu  s \ T'$, $\s{await} \ s$, or
$\s{watch}\ s \ 0$. In these cases the first component does
not increase while the size decreases.

\Proofitem{\bullet}
Suppose the program reduction is induced by the thread reduction:
\[
(C[(\s{thread}\ T)],E)  \act{\pset{T}} (C[0],E)~.
\]
Assume $\w{Call}(T)=(X,\ell)$ and $\w{Call}(C)=(Y,\ell')$.
By proposition \ref{call-cxt}, we have:
\[
\begin{array}{c}
\w{Call}(C[\s{thread}\ T]) = \w{Call}(\s{thread}\ T); \w{Call}(C) = 
(X,0); (Y,\ell') = (X\union Y,\ell') \\
\w{Call}(C[0]) = \w{Call}(0);\w{Call}(C) = (Y,\ell')~.
\end{array}
\]
Thus the first component does not increase while the size decreases.

\Proofitem{\bullet}
Finally, suppose the program reduction comes from the unfolding
of a recursive definition $A(\vc{x})= T$:
\[
C[A(\vc{s})] \act{\emptyset} C[[\vc{s}/\vc{x}]T]~.
\]
Assume  $\w{Call}(T)=(X,\ell)$ and $\w{Call}(C)=(Y,\ell')$.
Then
\[
\w{Call}(C[A(\vc{s})]) = (\pset{A}\union Y,\ell), \quad
\w{Call}(C[T]) = \w{Call}(T);\w{Call}(C) = (X,\ell);(Y,\ell')~.
\]
By hypothesis, $\pset{A}>X$.  We derive that
$\pset{A}\union Y >_{m,\w{Id}} X\union Y \geq_{m,\w{Id}} Y$,
and we notice that
$(X,\ell);(Y,\ell')$ equals $(X\union Y,\ell')$ if $\ell=0$
and $(X,\dcl)$, otherwise. \qed \\

Theorem \ref{reactivity-thm} provides a sufficient (but not necessary)
criteria to ensure reactivity. 

\begin{example}\label{example-reactivity}
Theorem \ref{reactivity-thm} provides a sufficient (but not necessary)
criteria to ensure reactivity.  Indeed, the precision of the analysis
can be improved by unfolding some recursive equations.
For instance, consider the thread $A$ defined by the system:
\[
\begin{array}{ll}
A &= (\s{watch} \ s_1 \ B);(\s{emit} \ s_4); A \\
B &= (\s{await} \ s_2);(\s{emit} \ s_3);\s{pause};B
\end{array}
\]
If we compute the corresponding $\w{Call}$ we obtain:
\[
\begin{array}{lll}
\w{Call}((\s{watch} \ s_1 \ B);(\s{emit} \ s_4); A) 
&= (\pset{B},0);(\emptyset,0);(\pset{A},0) 
&= (\pset{A,B},0)  \\

\w{Call}((\s{await} \ s_2);(\s{emit} \ s_3);\s{pause};B) 
&= (\emptyset,0);(\emptyset,0);(\emptyset,\dcl);(\pset{B},0) 
&=(\emptyset,\dcl)

\end{array}
\]
and obviously we cannot find a well founded order
such that $A>A$. However, if we unfold $B$ 
definition in $A$ then we obtain
$(\emptyset,\dcl);(\emptyset,0);(\pset{A},0) = 
(\emptyset,\dcl)$,
and the constraints are trivially satisfied.
\end{example}

\section{A tail-recursive model and a CPS translation}\label{cps-sec}
We introduce a more basic language of {\em tail recursive threads}
to which the `high level
language' introduced in section \ref{model-sec} can be compiled 
via a continuation passing style (CPS) translation. 
Tail recursive threads are denoted by
$t,t',\ldots$ and they are defined as follows
\[
t::= 0 \Alt A(\vc{s}) \Alt \s{emit} \ s.t \Alt
    \nu  s \ t \Alt \s{thread} \ t.t \Alt
    \s{present} \ s\  t \ \w{b} 
\]
where $A$ is a thread identifier with the usual conventions 
(cf. section \ref{model-sec}).
Let $b,b',\ldots$ stand for  {\em branching threads} defined as follows.
\[
\w{b}::= t \Alt \s{ite} \ s \ \w{b} \  \w{b}
\]
Branching threads can only occur in the `else'
branch of a \s{present} instruction and they are executed
only at the end of an instant once the presence or absence of
a signal has been established.
The small step thread reduction can be simply defined as follows:
\[
\begin{array}{llll}

(t_1)
&(\s{emit}\ s.t,E) 
&\act{\emptyset} (t,E[\true/s]) \\

(t_2)
&(\nu  s \ t,E) 
&\act{\emptyset} (t,E[\false/s]) 
&\mbox{if }s\notin \w{dom}(E)\\

(t_3)
&(A(\vc{s}), E) 
&\act{\emptyset}
([\vc{s}/\vc{x}]t,E)
&\mbox{if }A(\vc{x})=t \\

(t_4)
&(\s{present}\ s\ t\ b,E) 
&\act{\emptyset} (t,E)
&\mbox{if }E(s)=\true \\

(t_5)
&(\s{thread}\ t'.t, E)
&\act{\pset{t'}} (t,E) 

\end{array}
\]
The execution of the branching threads at the end of the instant
is defined as follows:
\[
\begin{array}{ll}

\lf 0 \rf_E = 0

&
\lf \s{present} \ s \ t \ b \rf_E =
\lc b \rc_E \\

\lc t \rc_E = t

&
\lc \s{ite}\ s\ b_1\ b_2 \rc_E = 
\left\{ 
\begin{array}{ll}
\lc b_1 \rc_E&\mbox{if }E(s)=\true \\
\lc b_2 \rc_E&\mbox{if }E(s)=\false 
\end{array}
\right.

\end{array}
\]
A program is now a finite non-empty multi-set of tail recursive
threads and program reduction is defined as in section
\ref{program-sec}.
We can define the instructions $\s{pause}$ and $\s{await}$ in `prefix
form' as follows:
\[
\begin{array}{ll}
\s{pause}.b  &= \nu  s \ \s{present} \ s \ 0 \ b \\
\s{await} \ s.t &=A, 
\quad\mbox{where: } A = \s{present} \ s \ t \ A, \quad
\set{\vc{s}}= \w{sig}(t)\union\set{s}~.
\end{array}
\]
Determinism is guaranteed by the design
of the language while reactivity can be enforced by a static analysis
similar (but simpler) than the one presented in section 
\ref{basic-properties-sec}. 

\subsection{CPS translation}
We denote with $\epsilon$ an empty sequence.
The translation $\dl \_ \dr$ described in table \ref{cps-translation} 
has 2 parameters:
(1) a thread $t$ which stands for the {\em default continuation} and
(2) a sequence  $\tau\equiv (s_1,t_1)\cdots (s_n,t_n)$.
If $s_i$ is the `first' (from left to right) signal which is present
then $t_i$ is the continuation.
Whenever we cross a \s{watch} statement we insert 
a pair $(s,t)$ in the sequence $\tau$.
Then we can translate the \s{await} statement 
with the \s{present} statement provided that at the
end of each instant we check (from left to right)
whether there is a pair $(s,t)$ in $\tau$ 
such that the signal $s$ is present. In this case,
the continuation $t$ must be run at the following instant.

Some later versions of the SL language include
a $(\s{when} \ s \ T)$ statement whose informal semantics is
to run $T$ (possibly over several instants) when $s$ is present.
It is possible to elaborate the CPS translation
to handle this operator. The idea is to introduce as an additional
parameter to the translation, the list of signals that have
to be present for the computation to progress.

In the translation of a thread identifier, say,
$A^{(t,\tau)}(\vc{x},\vc{s'})=\dl T \dr (t,\tau)$
the identifier $A^{(t,\tau)}$ takes as 
additional parameters the signal names free  in $(t,\tau)$.
For the sake of readability, in the following we will simply write
$A^{(t,\tau})(\vc{x})$ and omit the parameters $\vc{s'}$.

It is important to notice that 
the translation associates with an equation 
$A(\vc{x})=T$ a potentially infinite family of equations
$A^{(t,\tau)}(\vc{x})=\dl T \dr(t,\tau)$,
the index $(t,\tau)$ depending on the evaluation context.
However, whenever the evaluation contexts are `bounded'
in the sense described in the following section 
\ref{bounded-cxt},  only a finite number of indices are
needed and the CPS translation preserves the finiteness of the 
system of recursive equations.

\begin{example}\label{example-cps}
We compute the CPS translation of the thread $A$ in
example \ref{example-reactivity} (without unfolding). 
To keep the translation compact, we will use a slightly optimised 
CPS translation of the \s{pause} statement that goes as follows:
\[
\dl \s{pause} \dr(t,(s_1,t_1)\cdots(s_n,t_n)) = 
\s{pause}.\s{ite} \ s_1 \ t_1 (\cdots 
(\s{ite} \ s_n \ t_n \ t)\cdots )
\]
Then the translation can be written as follows:
\[
\begin{array}{llll}

A^{(0,\epsilon)} &= B^{(t_1,\tau_1)} 
&t_1              &= \s{emit} \ s_4.A^{(0,\epsilon)} \\
\tau_1           &= (s_1,t_1) 
&B^{(t_1,\tau_1)} &=\s{present} \ s_2 \ t_2 \ 
                  (\s{ite} \ s_1 \ t_1 \ B^{(t_1,\tau_1)}) \\
t_2     &=\s{emit} \ s_3.\s{pause}.\s{ite} \ s_1 \ t_1 \ B^{(t_1,\tau_1)}~.

\end{array}
\]
\end{example}

\begin{table}
\[
\begin{array}{ll}

\dl 0 \dr(t,\tau) 
&= t \\[+.3em] 

\dl T_1;T_2 \dr (t,\tau) 
&=  \dl T_1 \dr (\dl T_2 \dr(t,\tau), \tau) \\[+.3em]  

\dl \s{emit} \ s\dr (t,\tau)
&=\s{emit}\ s.t \\[+.3em] 

\dl \nu  s \ T \dr(t,\tau) 
&=\nu  s \ \dl T \dr(t,\tau),\quad\mbox{where: }s\notin \w{sig}(t)\union \w{sig}(\tau)  \\[+.3em]  

\dl \s{thread} \ T \dr(t,\tau) 
&=\s{thread} \ \dl T \dr(0,\epsilon).t   \\[+.3em] 


\dl \s{watch}\ s \ T \dr (t,\tau) 
&= \dl T \dr (t, \tau\cdot(s,t))  \\[+.3em] 

\dl \s{await}\ s \dr(t,\tau) 
&= \s{present} \ s\ t \ b, \quad\mbox{where: } 

\tau = (s_1,t_1)\cdots (s_m,t_m), \\


&\quad b\equiv (\s{ite} \ s_1 \ t_1 \ldots (\s{ite} \ s_m \ t_m \ A)\ldots),
\quad 
A = \s{present} \ s \ t \ b 

\\[+.3em]

\dl A(\vc{s}) \dr (t,\tau) 
& = A^{(t,\tau)}(\vc{s},\vc{s'}),
\quad
\mbox{where: }
\w{sig}(t,\tau)=\set{\vc{s'}}, \quad 
A(\vc{x}) = T, \\
&\qquad\qquad\qquad\quad
\set{\vc{x}}\inter \set{\vc{s'}}= \emptyset, \quad 
A^{(t,\tau)}(\vc{x},\vc{s'}) = \dl T \dr (t,\tau)~.
\end{array}
\]
\caption{A CPS translation}\label{cps-translation}
\end{table}

The translation is lifted to programs as follows:
$\dl P \dr = \pset{ \dl T \dr (0,\epsilon) \mid T \in P}$.
We show that a program generates exactly the same traces (cf. section 
\ref{trace-semantics}) as its CPS translation.
To this end, it is convenient to extend the CPS 
translation to evaluation contexts as follows:
\[
\begin{array}{ll}

\dl [~] \dr (t,\tau) &=(t,\tau) \\ 

\dl [~];T \dr (t,\tau) &=(\dl T \dr(t,\tau),\tau) \\

\dl \s{watch} \ s \ C \dr (t,\tau) 
&= \dl C \dr (t,\tau \cdot (s,t)) \\

\dl (\s{watch} \ s \ C);T \dr (t,\tau) 
&= \dl C \dr (\dl T \dr(t,\tau),\tau \cdot (s,\dl T \dr(t,\tau))) \\

\end{array}
\]
Then we note the following decomposition property of the CPS translation
whose proof is by induction on the evaluation context.

\begin{proposition}\label{cxt-cps}
For all $C$ evaluation context, $T$ thread, $t$ tail recursive thread, 
$\tau$ sequence,
\[
\dl C[T] \dr (t,\tau) = \dl T \dr(\dl C \dr (t,\tau))~.
\]
\end{proposition}


\begin{definition}
We define a relation $\cl{R}$ between threads in the source and
target language: $T \rel{R} t$ if either
(1) $t = \dl T \dr (0,\epsilon)$  or 
(2) $T=C[\s{await} \ s]$, $t= A$,
 and $A= \dl T \dr(0,\epsilon)$.
\end{definition}
  
The idea is that $T\rel{R} t$ if $t=\dl T \dr(0,\epsilon)$ up to the
unfolding of the recursive definition in the CPS translation of an
\s{await} statement.  The need for the unfolding arises when checking
the commutation of the CPS translation with the computation at the end
of the instant.  Then, we show that the relation $\cl{R}$ behaves as a
kind of weak bisimulation with respect to reduction and suspension and
that it is preserved by the computation at the end of the
instant. This point requires a series of technical lemmas which are
presented in appendix \ref{proof-cps-thm}.  In turn, these lemmas entail
directly the following theorem \ref{cps-thm}.

\begin{theorem}\label{cps-thm}
Let $P$ be a program. Then $\w{tr}(P) = \w{tr}(\dl P \dr)$.
\end{theorem}

\subsection{A static analysis to bound evaluation contexts}\label{bounded-cxt}
The source language allows an unlimited accumulation of evaluation
contexts.  To avoid stack overflow at run time, we define a simple
control flow analysis that guarantees that each thread has an
evaluation context of bounded size.  For instance, have this property:
(i) the fragment of the language using \s{loop} rather than recursive
definitions and (ii) programs where recursive calls under a \s{watch}
are guarded by a \s{thread} statement such as $A= (\s{watch} \ s \
\s{pause};(\s{thread} \ A))$.  On the other hand, fail this property
recursive definitions such as: (i) $A= \s{pause};A;B$ and (ii)
$A=(\s{watch} \ s\ \s{pause};A)$.

Let $L=\set{\epsilon,\kappa}$ be a set of labels. Intuitively,
$\epsilon$ indicates an empty evaluation context, while $\kappa$
indicates a (potentially) non-empty evaluation context.
Sequential composition and the \s{watch} statement increase the
size of the evaluation context while the \s{thread} statement 
resets its size to $0$. Following this intuition,
we define a function
$\w{Call}$ that associates with a thread and a label a set of pairs
of thread identifiers and labels.
\[
\begin{array}{c}
\w{Call}(0,\ell) = \w{Call}(\s{await} \ s,\ell) = \w{Call}(\s{emit} \
s,\ell)=\emptyset, 

\qquad \w{Call}(A,\ell) = \set{(A,\ell)}, \\[0.5em]

\w{Call}(\s{thread} \ T,\ell) = \w{Call}(T,\epsilon),

\qquad \w{Call}(T_1;T_2,\ell) = \w{Call}(T_1,\kappa)\union \w{Call}(T_2,\ell), \\[0.5em]

\w{Call}(\s{watch}\ s \ T,\ell) = \w{Call}(T,\kappa)~.

\end{array}
\]

\begin{definition}[constraints]
We denote with $\w{Cnst}(P)$ the least set of inequality and equality
constraints on thread identifiers such that for any equation 
$A(\vc{x})=T$ in the program $P$: (1) if $(B,\kappa)\in \w{Call}(T)$ then
$A>B\in \w{Cnst}(P)$ and (2) if $(B,\epsilon) \in \w{Call}(T)$ then
$A\geq B\in \w{Cnst}(P)$.
\end{definition}

If $\succeq$ is a pre-order we define:
(i) $x\simeq y$ if $x\succeq y$ and $y\succeq x$ and
(ii) $x\succ y$ if $x\succeq y$ and $x\not\simeq y$.

\begin{definition}[satisfaction]
We say that a pre-order $\succeq$ on thread identifiers satisfies the
constraints $\w{Cnst}(P)$ if: 
(1) $A>B \in \w{Cnst}(P)$ implies $A\succ B$,
(2) $A\geq B \in \w{Cnst}(P)$ implies $A\succeq B$, and 
(3) $\succ$ is well-founded.
\end{definition}

We can now state the correctness of our criteria whose proof
is delayed to appendix \ref{proof-bounded-cxt-prop}. The
reader may check the criteria on example
\ref{example-cps}.

\begin{proposition}\label{bounded-cxt-prop}
If there is a pre-order that satisfies $\w{Cnst}(P)$ then 
the CPS translation preserves the finiteness of the system of
equations.
\end{proposition}

\section{Expressivity}\label{expressivity-sec}
In this section we  present two basic results on the 
computational expressivity of the model. 
First, we show that reactive programs without signal
generation are  trace equivalent to {\em monotonic} 
deterministic finite state machines, modulo a natural encoding.
Second, we notice that the combination of recursion and signal name
generation allows to simulate the computation of two counter machines.
Thus, unlike the original SL language, it is not always possible to
compile our programs to finite state machines.

\subsection{Monotonic Mealy machines}
A {\em monotonic} Mealy machine is a particular Mealy
machine whose input and output alphabets are powersets and such that
the function that determines the output respects the inclusion order
on powersets. As for programs, we can associate with a 
monotonic Mealy machine a set of traces.

\begin{definition}[monotonic Mealy machine]\label{mealy-def}
A finite state, deterministic, reactive, and
monotonic Mealy machine (monotonic Mealy machine for short)
is a tuple $M=(Q,q_o,I,O,f_Q,$ $f_O)$ where 
$Q$ is a finite set of states, 
$q_o\in Q$ is the initial state,
$I=2^n$, $O=2^m$ for
$n,m$ natural numbers are the input and output alphabets,
respectively, $f_Q:I\times Q \arrow Q$ is the function
computing the next state, and 
$f_O:I\times Q \arrow O$ is the function computing the output
which is monotonic in the input, namely 
$X\subseteq Y$ implies $f_O(X,q)\subseteq f_O(Y,q)$.
\end{definition}

\begin{theorem}\label{monotonic1}
For every monotonic Mealy machine with 
input alphabet $I=2^n$ and output alphabet
$O=2^m$ there is a trace equivalent
program with $n$ input signals and $m$
output signals.
\end{theorem}
\Proof
The function $f_Q(\_,q)$ 
that for a given state $q$  computes the next state  
as a function of the input can be coded as a cascade of \s{ite}'s.
The function $f_O(\_,q)$ that for a given state $q$
computes the output as a function of the input can be coded
as the parallel composition of threads that emit a certain
output signal if a certain number of input signals is present
in the instant and do nothing otherwise. 

Next we develop some details.
Let $M=(Q,q_o,I,O,f_Q,f_O)$ with $I=2^n$ and $O=2^m$
be a monotonic Mealy machine.
We build the corresponding program.
We introduce signals $s_1,\ldots,s_n$ for the input and
signals $s'_{1},\ldots,s'_{m}$ for the output.
Moreover, we introduce a thread identifier $q$ for
every state $q\in Q$.
Given a state $q$, we associate with the function 
$f_Q(\_,q):2^n\arrow Q$ a branching thread $b(q)$.
For instance, if the function is defined by:
\[
\begin{array}{c}
f_Q((1,1),q)=q_1,
\quad
f_Q((1,0),q)=q_2,
\quad
f_Q((0,1),q)=q_3,
\quad
f_Q((0,0),q)=q_1,

\end{array}
\]
then the corresponding branching thread is:
\[
b(q) = \s{ite} \ s_1 \ (\s{ite}\ s_2 \ q_1 \ q_2) \ (\s{ite}\ s_2 \ q_3 \ q_1)
\]
For every state $q$, we introduce an equation of the
shape:
\begin{equation}\label{state-equation}
   q = \w{Output}(q).\s{pause}.b(q)
\end{equation}
where $\w{Output}(q)$ is intended to compute the output function
$f_O(\_,q): 2^n\arrow 2^m$. To formalise this,
we need some notation.
Let $X\subseteq \set{1,\ldots,n}$ denote an input symbol
and $j\in \set{1,\ldots,m}$. By monotonicity, if $X\subseteq Y$
and $j\in f_O(X,q)$ then $j\in f_O(Y,q)$.
Given a family of threads $\set{t_j}_{j\in J}$, we write
$\s{thread}_{j\in J} t_j.t$ for the thread that spawns,
in an arbitrary order, the threads $t_j$ and then runs $t$.
Given a set of input signals $\set{s_1,\ldots,s_k}$ 
and an output signal $s'_j$, we write 
$\s{await} \set{s_1,\ldots, s_k}.t$ for 
\[
\s{present} \ s_1 \ 
(\cdots (\s{present} \ s_k \ t \ 0 )\cdots) \ 0
\]
which executes $t$ in the first instant it is run 
if and only if all the signals  $s_1,\ldots,s_k$ are present,
and terminates otherwise. No signals are emitted in the instants
following the first one.
With these conventions $\w{Output}(q).t$ is an abbreviation for
\[
(\ \s{thread}_{X\subseteq \set{1,\ldots,n},\ j\in f_{O}(X,q)} \
(\s{await} \ \set{s_x \mid x\in X}.\ \s{emit} \ s'_j) \ ).\ t
\]
so that the explicit form for equation (\ref{state-equation}) is:
\[
q= (\ \s{thread}_{X\subseteq \set{1,\ldots,n}, \ j\in f_{O}(X,q) } \
 (\s{await} \ \set{s_x \mid x\in X}. \ \s{emit} \ s'_j) \ ).\ \s{pause}. \ b(q) ~.
\]
\qed

One may wonder whether our synchronous language may represent
{\em non-monotonic} Mealy machines. The answer to this question
is negative as long we adopt the encoding of the input above where
$2^n$ input symbols are mapped to $n$ signals. This fact
easily follows from the monotonicity property of the model
noted in section \ref{basic-properties-sec}. 
However, the answer is positive if we adopt a 
less compact representation where 
$n$ input symbols are mapped to $n$ signals.

Next we focus on the expressive power of the reactive programs 
we can write in the tail recursive calculus presented
in section \ref{cps-sec} {\em without signal generation} but with
general recursion and thread spawning.  

\begin{theorem}\label{monotonic2}
For every reactive tail recursive program 
with $n$ input signals and $m$
output signals and without signal generation
there is  a trace equivalent monotonic Mealy machine with
input alphabet $2^n$ and output alphabet $2^m$.
\end{theorem}
\Proof
The construction takes several steps but the basic idea is 
simple: it is useless to run twice or more times through
the same `control point' within the same instant. Instead
we record the set of control points that have been reached
along with the signals that have been emitted and in doing so
we are bound to reach a fixed point.

We start with some preliminary considerations that
allow to simplify the representation of programs.

\Proofitemb{(1)}
Since there is no signal generation a program
depends on a finite set $S_o$ of signal names.
As a first step we can remove parameters from
recursive equations. To this end, replace every
parametric equation $A(\vc{x})=t$ with a finite number
of equations (without parameters) of the shape $A_\vc{s} =
[\vc{s}/\vc{x}]t$ for $\vc{s}$ ranging over tuples of signal names in
$S_o$.

\Proofitemb{(2)}
As a second step, we put the recursive equations in normal form.
By introducing auxiliary thread identifiers, we may assume the
equations have the shape $A=t$ where 
\[
\begin{array}{ll}
t &::= 0 \Alt \s{emit}\ s.B \Alt \s{present} \ s\ B \ b \Alt 
\s{thread} \ B.B'  \\
b &::= A \Alt \s{ite} \ s\ b \ b
\end{array}
\]
We denote with $\w{Id}_o$ the finite set of thread identifiers.

\Proofitemb{(3)}
Because there is no signal name generation, we may simply represent
the environment $E$ as a subset of $S_o$ and because the threads
are in normal form we may simply represent a program $P$
as a multi-set of identifiers in $\w{Id}_o$.
The small step reduction of the pair $(P,E)$ is then
described as follows:
\[
(P\union \pset{A},E) \arrow
\left \{
\begin{array}{ll}
(P\union \pset{B},E\union\set{s}) &\mbox{if }A=\s{emit} \ s.B \\
(P\union \pset{B},E) &\mbox{if }A=\s{present} \ s \ B\  b,\ s\in E \\
(P\union \pset{B_1,B_2},E) &\mbox{if }A=\s{thread} \ B_1.B_2 
\end{array}\right.
\]
Notice that in this presentation, the unfolding of recursive
definitions is kept implicit.
If the program is reactive we know that the evaluation
of a pair $(P,E)$ eventually terminates in a configuration
$(P',E')$ such that if $A\in P'$ then either $A=0$ or
$A=\s{present} \ s \ B\ b$ and $s\notin E'$. The evaluation
at the end of the instant $\lf P' \rf_{E'}$ is then a particular
case of the one defined in section \ref{cps-sec} for tail recursive threads
and produces again a multi-set of thread identifiers.

\Proofitemb{(4)} We now consider an alternative representation
of a program as a {\em set} $q$ of identifiers in $\w{Id}_o$. 
We define a small step reduction on configurations $(q,E)$ as
follows:
\[
(q\union \set{A},E) \arrow
\left \{
\begin{array}{ll}
(q\union \set{A,B},E\union\set{s}) 
&\mbox{if }A=\s{emit} \ s.B, \ (B\notin q\union\set{A} \mbox{ or } s\notin E) \\

(q\union \set{A,B},E) 
&\mbox{if }A=\s{present} \ s \ B\  b, \ s\in E, \ B\notin q\union \set{A}  \\

(q\union \set{A,B_1,B_2},E) &\mbox{if }A=\s{thread} \ B_1.B_2,
\ \set{B_1,B_2}\not\subseteq q\union\set{A}
\end{array}\right.
\]
Note that at each reduction step either the program $q$ or the 
environment $E$ increase strictly while the other component does not decrease.
Consequently, this reduction process (unlike the previous one)
necessarily terminates. The evaluation at the end of the instant
is now defined as follows:
\[
\lf q \rf_E =\set{A\in q \mid A=0} \union
\set{\lc b \rc_E \mid A\in q, A=\s{present} \ s  \ B \ b,\mbox{ and } s\notin E}~.
\]
Notice that $q$ may contain, {\em e.g.}, a thread identifier 
$A$ such as $A=\s{emit} \ s.B$ and
that $A$ is removed by the function $\lf\_\rf_E$.

\Proofitemb{(5)}
We now relate the two representations of the programs and
the associated evaluation strategies where 
if $P$ is a multi-set we let
$\w{set}(P)=\set{A\mid A\in P}$ be the corresponding 
set where we forget multiplicities.

\begin{lemma}\label{monotonic-lemma}
Suppose  $(P_1,E_1)\arrow \cdots \arrow (P_n,E_n)$ with
$n\geq 1$ and $q=\w{set}(P_1\union\cdots \union P_n)$.
Then:

\Defitemf{(1)}
If $(P_n,E_n)\arrow (P_{n+1},E_{n+1})$ 
then
either $E_{n}=E_{n+1}$ and $\w{set}(P_{n+1})\subseteq q$ or
$(q,E_n)\arrow (q',E_{n+1})$ and 
$q'=\w{set}(P_1\union\cdots \union P_{n+1})$.

\Defitem{(2)}
If $(q,E_n)\arrow (q',E_{n+1})$ then
$(P_n,E_n) \arrow (P_{n+1},E_{n+1})$ and
$q'=\w{set}(P_1\union\cdots \union P_{n+1})$.

\Defitem{(3)}
If $(P_n,E_n)\susp$ then
$\w{set}(\lf P_n \rf_{E_{n}}) = \lf q \rf_{E_{n}}$.
\end{lemma}
\Proof
\Proofitemf{(1)} 
By case analysis on the small step reduction for multi-sets. 

\Proofitem{(2)}
By case analysis on the small step reduction for sets.  Note that if
the reduction rule is applied to $A\in q$ then necessarily $A\in
P_n$. Indeed, if $A\in P_k$ and $A\notin P_{k+1}$ with $k<n$ we can
conclude that a reduction rule has been applied to $A$ on the
multi-set side and this contradicts the hypotheses for the firing of
the rule on the set side.

\Proofitem{(3)}
We check that if $A=0$ and $A\in q$ then $A\in P_n$ and
that if $A=\s{present} \ s\ B \ t$, $s \notin E_n$ and $A\in q$ then
$A\in P_n$. \qed 

\Proofitemb{(6)}
We define 
\[
\w{Closure}(q,E)=(q',E')
\mbox{ if }
(q,E)\arrow \cdots \arrow (q'',E')\not\arrow
\mbox{ and }
q'=\lf q ''\rf_{E'}
\]
The \w{Closure} operator is well defined because the
reduction relation is strongly confluent and it 
always terminates.

\Proofitemb{(7)}
As a final step, 
given a reactive program $P$ in normal form with identifiers 
$\w{Id}_o$, $n$ input signals
$s_1,\ldots,s_n$ and $m$ output signals $s'_1,\ldots,s'_m$,
we build a trace equivalent monotonic Mealy machine 
$M=(Q,q_o,I,O,f_Q,f_O)$ as follows: $Q=2^{\w{Id}_o}$,
$q_o=\w{set}(P)$, $I=2^{n}$, $O=2^{m}$, and 
$(f_Q(E,q),f_O(E,q))=\w{Closure}(q,E)$. \qed \\


By combining theorems \ref{monotonic1} and \ref{monotonic2}, 
we can conclude that the reactive programs we can write 
without signal generation are exactly those
definable by monotonic Mealy machines modulo a natural encoding.

\subsection{Undecidability}
The following result 
can be used to show that various questions
about the behaviours of programs are undecidable.
The encoding idea is similar to the one presented
for CCS in \cite{Milner89}. The details are presented
in appendix \ref{proof-undecidability-thm}.

\begin{theorem}\label{undecidability-thm}
For any deterministic 2-counter machine there is 
a reactive program with signal generation 
that will eventually emit on a certain signal 
if and only if the computation
of the 2-counter machine terminates.
\end{theorem}

\section{Program equivalence}\label{equiv-sec}
The formalisation of the SL model we have considered so far is close
to an abstract machine. Typical symptoms include an {\em ad hoc}
definition of $\alpha$-renaming (cf. section
\ref{basic-properties-sec}), a global notion of environment, and the
fact that roughly threads compose but do not reduce while programs
reduce but do not compose.  We introduce next an alternative
description of the tail recursive model featuring a uniform notation
for threads, programs, and environments. This alternative description
is instrumental to the development of a notion of program equivalence
based on the concept of bisimulation following a CCS style.  
The theory is built so that it does not depend on the determinacy of
the language.  Indeed {\em practical} extensions of the language have
been considered where signals carry data values and the act of
receiving a value may introduce non-determinism. A theory of program
equivalence should be sufficiently robust to accommodate these
extensions.

\subsection{Programs}
We extend the syntax of tail recursive threads so that
it includes both environments and programs in a uniform notation.
\[
\begin{array}{ll}
P &::= 0 \Alt \s{emit} \ s \Alt 
     \s{present} \ s\  P \ B \Alt 
     P\mid P \Alt 
     \nu  s \ P \Alt 
     A(\vc{s})  \\ 

B &::= P \Alt \s{ite} \ s\ B \ B
\end{array}
\]
We refrain from introducing syntax like 
$`\s{emit} \ s.P$ and $`\s{thread} \ P'.P$
which can be understood as syntactic sugar for 
$(\s{emit} \ s) \mid P$ and $P'\mid P$, respectively.

\subsection{Actions and labelled transition system}
Actions are denoted by $\alpha, \alpha',\ldots$ and they 
are defined by the grammar:
$\alpha::= \tau \Alt s \Alt \ol{s}$.
We write $s\in \alpha$ if $\alpha=s$ or $\alpha=\ol{s}$.
We define a {\em labelled transition system} which is similar
to the one for CCS except for a different treatment of
emission which is {\em persistent} within an instant.
Technically, (i) an emission behaves as a replicated output 
(rule $(\w{out})$) and (ii) in the continuation of a \s{present} statement
the tested signal is still emitted  (rule $(\w{in})$); this
guarantees that the continuation can only evolve in an environment
where the signal $s$ is emitted.\footnote{This is close in spirit, 
if not in the technical
development, to Prasad's Calculus of Broadcasting Systems \cite{P95};
see also \cite{HR98}. }
\[
\begin{array}{lclc}

(\w{out})
& \infer{}{\s{emit} \ s \act{\ol{s}} \s{emit} \ s} 

&(\w{in})
&
\infer{}{\s{present} \ s\  P \ \w{B} \act{s} P \mid (\s{emit} \ s)} \\[+.5em]

(\tau)
& \infer{P_1 \act{s} P'_1\quad P_2 \act{\ol{s}} P'_2}
{P_1\mid P_2 \act{\tau} P'_1 \mid P'_2}

&(\w{par})
&
\infer{P_1\act{\alpha} P'_1}{P_1 \mid P_2 \act{\alpha} P'_1\mid P_2}
\\[+.5em]

(\nu)
&
\infer{P\act{\alpha} P'\quad s\notin \alpha}
{\nu  s\ P \act{\alpha} \nu  s \ P'} 

&(\w{rec})
&
\infer{A(\vc{x})=P}
{A(\vc{s})\act{\tau} [\vc{s}/\vc{x}]P}

\end{array}
\]
As usual, we omit the symmetric rules for $(\w{par},\tau)$.
We note the following properties of the labelled transition
system where $=$ stands for syntactic identity up to renaming of bound
names.

\begin{proposition}\label{lts-prop}
\Defitemf{(1)}
If $P\act{\ol{s}} P'$ then $P= P'$.


\Defitem{(2)}
If $P\act{\ol{s}} P$ and $P\act{\alpha}P'$ then
$P'\act{\ol{s}}P'$.

\Defitem{(3)}
If $P\act{s} P'$  then $P'\act{\ol{s}}P'$.
\end{proposition}

\subsection{End of the instant}
We define the computation at the end of the instant
while relying on the following notation:
$P\act{\alpha} \cdot$  for $\xst{P'}{P\act{\alpha} P'}$ and
$P\susp$ for $\neg(P\act{\tau} \cdot)$.
Suppose $P\susp$ and all bound signal names in $P$
are renamed so as to be distinct and different from the free signal names.
First, we compute the set of emitted signals $S=\w{Em}(P)$ as follows:
\[
\begin{array}{c}
\w{Em}(\s{emit} \ s)=\set{s},
\quad
\w{Em}(0)=\w{Em}(\s{present} \ s\  P \ B)=\emptyset, \\[+.5em]

\w{Em}(P_1\mid P_2) = \w{Em}(P_1)\union \w{Em}(P_2),
\quad
\w{Em}(\nu  s \ P)=\w{Em}(P)~.
\end{array}
\]
Second, we compute
$\lf P \rf  = \lf P \rf_{\w{Em}(P)}$
where we remove all emitted signals and compute the $B$ branches
relying on the auxiliary functions $\lf\_ \rf_S$ and 
$\lc \_ \rc_S$ defined as follows:
\[
\begin{array}{c}
\lf \s{emit} \ s \rf_S =  \lf 0 \rf_S = 0,
\quad
\lf \s{present} \ s\  P \ B \rf_S = \lc B \rc_S,  \\[+.5em]

\lf \nu  s \ P \rf_S =\nu  s \  \lf P \rf_S,
\quad
\lf P_1\mid P_2 \rf_S = \lf P_1 \rf_S \mid \lf P_2 \rf_S, \\[+.5em]

\lc P \rc_S = P,
\quad
\lc \s{ite}\ s\ B_1\ B_2 \rc_S = 
\left\{ 
\begin{array}{ll}
\lc B_1 \rc_S&\mbox{if }s\in S \\
\lc B_2 \rc_S&\mbox{if }s\notin S~. 
\end{array}
\right.
\end{array}
\]
One can verify that the function $\lf \_ \rf$ 
is invariant under $\alpha$-renaming:
if $P_1=P_2$ then $\lf P_1 \rf = \lf P_2 \rf$.

\subsection{Barbed and contextual bisimulations}
As usual, we write  $P\wact{\tau} P'$ for $P (\act{\tau})^* P'$ and
$P\wact{\alpha} P'$ with $\alpha\neq \tau$ for
$P (\wact{\tau})(\act{\alpha})(\wact{\tau}) P'$.

\begin{definition}\label{wl-susp-def}
We define: 
\[
\begin{array}{lll}
P \wsusp &\mbox{if } \xst{P'}{P\wact{\tau} P' \mand P'\susp}
&\mbox{(weak suspension)} \\ 
P\lsusp 
&\mbox{if } P\act{\alpha_{1}} P_1 \cdots \act{\alpha_{n}} P_n,
\quad n\geq 0, \mand P_n\susp 
&\mbox{(L-suspension)}
\end{array}
\]
\end{definition}

Obviously $P\susp$ implies $P\wsusp$ which in turn implies 
$P\lsusp$. The L-suspension predicate (L for labelled) plays
an important role in the following definitions of bisimulation.

\begin{definition}
A (static) context $C$ is defined by
$C::= [~] \Alt C\mid P \Alt \nu s \ C$.
\end{definition}

\begin{proposition}\label{char-lsusp-prop}
Let $P$ be a program. The following are equivalent:

\Defitem{(1)} $P \lsusp$.

\Defitem{(2)} There is a program $Q$ such that $(P\mid Q)\wsusp$.

\Defitem{(3)} There is a static context $C$ such that
$C[P] \lsusp$.
\end{proposition}
\Proof
\Proofitemf{(1\Arrow 2)}
Suppose $P_0 \act{\alpha_{1}} P_1 \cdots \act{\alpha_{n}} P_n$ and
$P_n\susp$.
We build $Q$ by induction on $n$.
If $n=0$ we take $Q=0$. Otherwise, suppose $n>0$.
By inductive hypothesis, there is $Q_1$ such that
$(P_1 \mid Q_1)\wsusp$. We proceed by case analysis on the
first action $\alpha_1$. We may assume $\alpha_1$ is
not an emission action for otherwise we can build a
shorter sequence of transitions.

\Defitem{(\alpha_1=\tau)} Then we take $Q=Q_1$ and
$(P_0\mid Q_1)\act{\tau} (P_1 \mid Q_1)$.

\Defitem{(\alpha_1=s)} 
Let $Q= (Q_1 \mid \ol{s})$. We have
$(P_0\mid Q) \act{\tau} (P_1 \mid Q_1 \mid \ol{s})$.
Since $P_1 \act{\ol{s}} P_1$, 
we observe that $(P_1\mid Q_1) \wsusp$ implies
$(P_1 \mid Q_1 \mid \ol{s}) \wsusp$.

\Proofitem{(2\Arrow 3)}
Take $C=[~]\mid Q$.

\Proofitem{(3\Arrow 1)}
First, check by induction on a static context $C$
that $P\act{\tau}\cdot$ implies
$C[P]\act{\tau} \cdot$. Hence $C[P]\susp$ implies
$P\susp$.
Second, show that $C[P]\act{\alpha} Q$ implies
that $Q=C'[P']$ and either $P=P'$ or  $P\act{\alpha'}P$.
Third, suppose 
$C[P] \act{\alpha_{1}} Q_1 \cdots \act{\alpha_{n}} Q_n$
with $Q_n\susp$. Show by induction on $n$ that
$P\lsusp$. Proceed by case analysis on the context $C$ and
the action $\alpha_1$. \qed \\

Interestingly, the second characterisation, shows that the
L-suspension predicate can be defined just in terms of the $\tau$
transitions and the suspension predicate. This means that the
following definitions of barbed and contextual bisimulation can be given
{\em independently} of the labelled transition system.

\begin{definition}[barbed bisimulation]
A symmetric relation $R$ on programs is a barbed bisimulation if
whenever $P\rl{R} Q$ the following holds:

\Defitem{(B1)}
If $P\act{\tau} P'$ then
$\xst{Q'}{Q\wact{\tau} Q' \mand P'\rl{R}Q'}$.

\Defitem{(B2)}
If $P\susp$ then 
$\xst{Q'}{Q\wact{\tau} Q', Q'\susp, P \rl{R}Q',\mand 
\lf P \rf \rl{R} \lf Q' \rf}$.

\Defitem{(B3)}
If $P\act{\ol{s}}\cdot$ and $P\lsusp$ then
$\xst{Q'}{Q\wact{\tau} Q', Q'\act{\ol{s}}\cdot, \mand P \rl{R}Q'}$.

\smallskip\noindent
We denote with $\bbis$ the largest barbed bisimulation.
\end{definition}

It is easily checked that $\bbis$ is reflexive and transitive.
A reasonable notion of program equivalence should be preserved
by the static contexts. We define accordingly a notion of contextual
bisimulation.\footnote{Here we adopt the notion of contextual equivalence
introduced by \cite{HY95} for the $\pi$-calculus. An alternative approach
is to consider a notion of {\em barbed equivalence} \cite{MS92}. 
We refer to \cite{FG98} for a comparison of the two methods.}

\begin{definition}[contextual bisimulation]
A symmetric relation $R$ on programs is a contextual bisimulation if
it is a barbed bisimulation (conditions B1-3) and moreover 
whenever $P\rl{R} Q$ then

\Defitem{(C1)} $C[P] \rl{R} C[Q]$, for any context $C$.

\smallskip\noindent
We denote with $\cbis$ the largest contextual bisimulation.
\end{definition}

Again it is easily checked that $\cbis$ is reflexive and transitive.
By its very definition, it follows that $P\cbis Q$ implies $C[P]\cbis
C[Q]$ and $P\bbis Q$.

\subsection{Labelled bisimulation}\label{lbis-sec}
Aiming at a more effective description of the notion
of contextual bisimulation, we introduce a notion
of {\em labelled} bisimulation.

\begin{definition}[labelled bisimulation]
A symmetric relation $R$ on programs is a labelled bisimulation
if it is a barbed bisimulation (conditions B1-3) and moreover whenever 
$P\rl{R} Q$ the following holds:

\Defitem{(L1)}
If $P'= (P\mid S) \susp$ with 
$S= \s{emit} \ s_1 \mid \cdots \mid \s{emit} \ s_n$,
$n\geq 0$ 
then
$\xst{Q'}{(Q \mid S) \wact{\tau} Q',\  
Q'\susp,\  P'\rl{R} Q', \ \mand \lf P'\rf \rl{R} \lf Q'\rf}$.

\Defitem{(L2)}
If $P\act{s} P'$ then either 
$\xst{Q'}{(\ Q\wact{s} Q'\mand P'\rl{R} Q')}$ or
$\xst{Q'}{( \ Q\wact{\tau} Q'\mand 
P'\rl{R} (Q'\mid \s{emit} \ s) \ )}$.

\smallskip\noindent
We denote with $\lbis$ the largest labelled bisimulation.
\end{definition}

\begin{remark}
\Defitemf{(1)}
Condition $(L1)$ strengthens $(B2)$ therefore in the following proof
the analysis of $(B2)$ is subsumed by the one of $(L1)$.
To see the necessity of condition $(L1)$, consider
\[
P=\s{present} \ s_1 \ 0 \ (\s{ite} \ s_2 \ (\s{emit} \ s_3) \ 0) 
\quad \mand
\quad
Q=\s{present} \ s_2 \ 0 \  0~.
\]
Then $P\susp$, $Q\susp$, and
$\lf P \rf = \lf Q \rf = 0$ so that 
conditions $(B1-3)$ and $(L2)$ are satisfied. 
However, if we plug $P$ and $Q$ in the context 
$[~] \mid (\s{emit} \ s_2)$
then the resulting programs exhibit different behaviours.
It is not difficult to show that condition $(L1)$ can be optimised so
that we only consider emissions on signals which are free in the programs
under consideration. For instance, a simple 
corollary of this optimisation is that
labelled bisimulation is decidable for programs {\em without}
recursive definitions.

\Defitem{(2)}
Condition $(L2)$ has already 
appeared in the literature in the context of the 
asynchronous $\pi$-calculus \cite{ACS98}.

\Defitem{(3)}
There is no condition for the emission because 
by proposition \ref{lts-prop}
condition $(B3)$
is equivalent to the following one: 
if $P\act{\ol{s}}P'$ and $P'\lsusp$ then
$\xst{Q'}{(\ Q\wact{\ol{s}}Q' \mand P'\rl{R}Q' \ )}$.

\Defitem{(4)}
The condition $P\lsusp$ in  $(B3)$ is always
satisfied by reactive programs which are those we are really
interested in.  We will see in section \ref{bis-trace-sec}, 
that thanks to strong confluence,
the condition $P\lsusp$ can be replaced by the condition  
$P\wsusp$ or equivalently by the condition $P\susp$.
However, one should keep in mind that there are 
non-deterministic extensions of the language where this 
identification fails and where moreover the definitions
based on the weaker conditions $P\susp$ or $P\wsusp$ 
lead to notions of labelled bisimulation which are not
preserved by parallel composition. For this reason,
our definitions of bisimulation are based on the L-suspension
predicate.
\end{remark}

We can now state the main result of this section.

\begin{theorem}\label{lab-bis-char}
$P \cbis Q$ iff $P\lbis Q$.
\end{theorem}

We outline the proof argument which is developed in the following.
First, we note that labelled bisimulation equates all programs which cannot
L-suspend and moreover it never equates a program which 
L-suspends to one which cannot.
Second,
we introduce a notion of {\em strong} labelled bisimulation which is
contained in labelled bisimulation.  It is shown that strong labelled
bisimulation satisfies some useful laws like associativity,
commutativity, commutation of signal name generation, $\ldots$ 
Third,
we develop a notion of labelled bisimulation up to strong labelled
bisimulation that considerably simplifies reasoning about labelled
bisimulation.  
Fourth, we show that $\cbis$ is a
labelled bisimulation up to strong labelled bisimulation 
so that $P\cbis Q$ implies $P\lbis Q$.
Fifth, we show that labelled bisimulation is preserved
by parallel composition with signal emission, it is reflexive and
transitive, and it is preserved by signal name generation, parallel
composition, and the \s{present} operator.  In particular, it follows
that $\lbis$ is preserved by the static contexts, {\em i.e.}, $\lbis$ is a
contextual barbed bisimulation and therefore $P\lbis Q$ implies
$P\cbis Q$.

\subsection{Labelled bisimulation and L-suspension}
We observe some remarkable properties of the L-suspension predicate.

\begin{proposition}\label{l-suspension}

\Defitem{(1)}
If $\neg P \lsusp$ and $\neg Q \lsusp$ then
$P\lbis Q$.

\Defitem{(2)}
If $P \lbis Q$ and $P\lsusp$ then $Q\lsusp$.
\end{proposition}

\Proof First we note the following properties:

\Defitem{(A)} By proposition \ref{char-lsusp-prop},
if  $(P\mid Q)\lsusp$ then $P\lsusp$.

\Defitem{(B)} By definition, if $\neg P\lsusp$ and $P\act{\alpha} P'$ then
$\neg P'\lsusp$.

\Proofitem{(1)}
We show that $\set{(P,Q) \mid \neg P\lsusp \mand \neg Q\lsusp}$
is a labelled bisimulation.

\Defitem{(B1)}
By (B), if $\neg P\lsusp$ and $P\act{\tau} P'$ then
$\neg P'\lsusp$.


\Defitem{(B3)} The hypothesis is not satisfied.

\Defitem{(L1)} By (A), if $\neg P \lsusp$
then $\neg (P\mid S) \lsusp$. Hence $\neg (P\mid S) \susp$.

\Defitem{(L2)} By (B), if  $\neg P\lsusp$ and $P\act{s} P'$ then
$\neg P'\lsusp$. Then we match the transition with
$Q\wact{\tau} Q$ and by (A) 
$\neg Q \lsusp$ implies $\neg (Q\mid (\s{emit} \ s))\lsusp$. 

\Proofitem{(2)} We proceed by induction on the shortest reduction
such that $P\act{\alpha_{1}} P_1 \cdots \act{\alpha_{n}} P_n$ and
$P_n \susp$.
Note that in such a reduction no emission action $\ol{s}$ occurs
(otherwise a shortest reduction can be found).
If $n=0$ then $(B2)$ requires $Q\wact{\tau} Q'$ and $Q'\susp$.
Hence $Q\lsusp$.
If $n>0$ then we consider the first action $\alpha_1$.
If $\alpha_1 =\tau$ then $(B1)$ requires 
$Q\wact{\tau} Q_1$ and $P_1\lbis Q_1$.
Then $Q_1 \lsusp$ by inductive hypothesis on $P_1$.
Hence $Q\lsusp$.
If $\alpha_1 = s$ then we have to consider two cases.
If $Q\wact{s} Q_1$ and $P_1\lbis Q_1$ then $Q_1\lsusp$
by inductive hypothesis on $P_1$. Hence $Q\lsusp$.
If on the other hand $Q\wact{\tau} Q_1$ and 
$P_1\lbis Q_1\mid (\s{emit} \ s)$ then
$Q_1\mid (\s{emit} \ s)\lsusp$. Hence by (A) $Q_1\lsusp$,
and $Q\lsusp$. \qed

\subsection{Strong labelled bisimulation and an up-to technique}
To bootstrap reasoning about labelled bisimulation,
it is convenient to introduce a much stronger notion of labelled
bisimulation.

\begin{definition}[strong labelled bisimulation]\label{slb-def}
A symmetric relation $R$ on programs is a strong 
labelled bisimulation if whenever $P\rl{R} Q$ the following holds:

\Defitem{(S1)} $P\act{\alpha} P'$ implies
$\xst{Q'}{Q\act{\alpha} Q' \mand P'\rl{R} Q'}$.

\Defitem{(S2)} $(P\mid S) \susp$ with 
$S=(\s{emit} \ s_1) \mid \cdots \mid (\s{emit} \ s_n)$, $n\geq 0$ implies 
$(P \mid S) \rl{R} (Q \mid S)$ and
$\lf P \mid S \rf \rl{R} \lf Q \mid S \rf$.\footnote{The condition $(Q\mid S)\susp$ follows by $(S1)$.}

\smallskip\noindent
We denote with $\sbis$ the largest strong labelled bisimulation.
\end{definition}

Note that in definition \ref{slb-def} not only we forbid
weak internal moves but we also drop the convergence condition
in $(B3)$ and the possibility of matching an input with
an internal transition in $(L2)$. For this reason, we adopt
the notation $\sbis$ rather than the usual $\sim_L$.
We say that a relation $R$ is a strong labelled bisimulation up to
strong labelled bisimulation if the conditions $(S1-2)$ hold when
we replace $R$ with the larger relation $(\sbis) \comp R \comp (\sbis)$.
Strong labelled bisimulation enjoys some useful properties whose
standard proof is delayed to appendix \ref{proof-lemma-sbis}

\begin{lemma}\label{lemma-sbis}
\Defitemf{(1)}
$\sbis$ is a reflexive and transitive relation.

\Defitem{(2)}
If $P\sbis Q$ then $P\lbis Q$.

\Defitem{(3)} The following laws hold:
\[
\begin{array}{ll}
P\mid 0 \sbis P,

&P_1 \mid (P_2 \mid P_3) \sbis (P_1\mid P_2) \mid P_3,  \\

P_1 \mid P_2 \sbis P_2 \mid P_1, 
&\nu s \ P_1 \mid P_2 \sbis \nu s\ (P_1 \mid P_2)
\mbox{ if }s\notin \w{sig}(P_2).
\end{array}
\]
\Defitemf{(4)}
If $P\sbis Q$ then $P\mid S \sbis Q\mid S$ where
$S=P_1 \mid \cdots \mid  P_n$ and 
$P_i=0$ or $P_i=(\s{emit} \ s_i)$, for $i=1,\ldots,n$, $n\geq 0$.

\Defitem{(5)}
If $R$ is a strong labelled bisimulation up to strong labelled bisimulation
then $(\sbis) \comp R \comp (\sbis)$ is a strong labelled bisimulation.

\Defitem{(6)}
If $P\act{\ol{s}}\cdot$ then 
$P\sbis P\mid (\s{emit} \ s)$.

\Defitem{(7)}
If $P_1\sbis P_2$, 
then $\nu s \ P_1 \sbis \nu s \ P_2$ and 
$P_1\mid Q  \sbis P_2 \mid Q$.
\end{lemma}

We use strong labelled bisimulation in the context of 
a rather standard `up to technique'.

\begin{definition}
A relation $R$ is a labelled bisimulation up to $\sbis$ if 
the conditions $(B1-3)$ and $(L1-2)$ are satisfied when 
replacing the relation $R$ with the (larger) relation 
$(\sbis)\comp R \comp (\sbis)$.
\end{definition}

\begin{lemma}
Let $R$ be a labelled bisimulation up to $\sbis$. Then:

\Defitem{(1)} 
The relation $(\sbis)\comp R \comp (\sbis)$ is a labelled
bisimulation.

\Defitem{(2)} If $P\rl{R} Q$ then $P\lbis Q$.
\end{lemma}
\Proof
\Proofitemf{(1)}
A direct diagram chasing using the congruence properties
of $\sbis$.

\Proofitem{(2)}
Follows directly from (1). \qed \\

\subsection{Characterisation}
As a first application of the `up to technique',
we show that $P\cbis Q$ implies $P \lbis Q$.

\begin{lemma}\label{cbis-lbis}
$\cbis$ is a labelled bisimulation up to $\sbis$.
\end{lemma}
\Proof  Suppose $P\cbis Q$. We check conditions $(L1-2)$.

\Proofitem{(L1)} Suppose $S=(\s{emit} \ s_1)\mid \cdots \mid (\s{emit} \ s_n)$
and $(P\mid S) \susp$.
Since $\cbis$ is preserved by parallel composition we derive
$P\mid S \cbis Q\mid S$. Then we
conclude by applying condition $(B2)$.

\Proofitem{(L2)} Suppose $P\act{s}P'$. By lemma \ref{lemma-sbis}(6), this implies
$P'\sbis P'\mid (\s{emit} \ s)$. Since $\cbis$ is preserved by
parallel composition we know $P\mid (\s{emit} \ s ) \cbis 
Q\mid (\s{emit} \ s)$.
From this and the fact that
$P\mid (\s{emit} \ s) \act{\tau} P'\mid (\s{emit} \ s)$
condition $(B1)$ allows to derive that 
$Q\mid (\s{emit} \ s) \wact{\tau} Q'\mid (\s{emit} \ s)$ and 
$P'\mid (\s{emit} \ s) \cbis Q'\mid (\s{emit} \ s)$.
Two cases may arise: 
(1) $Q\wact{s} Q'$. Then we have 
$P'\sbis P'\mid (\s{emit} \ s) \cbis Q'\mid (\s{emit} \ s) \sbis 
Q'$.
(2) $Q\wact{\tau} Q'$. Then we have 
$P'\sbis P'\mid (\s{emit} \ s) \cbis Q'\mid (\s{emit} \ s)$.
In both cases we close the diagram up to $\sbis$. \qed  \\

As a second application of the `up to technique' we 
prove some desirable congruence properties of the labelled
bisimulation (the proofs are delayed to appendix 
\ref{proof-lemma-bis-congr}).
Assume  $\s{pause}.B$ abbreviates $\nu s \ \s{present} \ s \ 0 \ B$ for
$s\notin \w{sig}(B)$.
We write $B_1 \lbis B_2$ if  $\s{pause}.B_1 \lbis \s{pause}.B_2$.

\begin{lemma}\label{lemma-bis-congr}
\Defitemf{(1)}
If $P\lbis Q$ then $P\mid (\s{emit} \ s) \lbis Q \mid (\s{emit} \ s)$.

\Defitem{(2)}
The relation $\lbis$ is reflexive and transitive.

\Defitem{(3)}
If $P\lbis Q$ then $\nu s\ P \lbis \nu s \ Q$.


\Defitem{(4)}
If $P_1 \lbis P_2$ then 
$P_1\mid Q \lbis P_2 \mid  Q$.

\Defitem{(5)}
If $P\lbis P'$ and $B\lbis B'$ then
$\s{present} \ s\ P \ B \lbis 
\s{present} \ s\ P' \ B'$.
\end{lemma}

The lemma above entails that  $\lbis$ is preserved by
static contexts. Hence $P \lbis Q$ implies $P\cbis Q$.
This remark combined with lemma \ref{cbis-lbis} concludes the proof
of theorem \ref{lab-bis-char}.

\subsection{Exploiting confluence}\label{bis-trace-sec}
We can easily adapt the trace semantics 
presented in section \ref{trace-semantics} to the present context.
If $P$ is a program we write  ($\Pi$ for the parallel composition): 
\[
P\act{I/O} P'\mbox{ if } 
P\mid P_I\wact{\tau} P'',\mbox{ with }P_I = \Pi_{s\in I} \ol{s}, 
\quad P''\susp, \quad O=\set{s\mid P''\act{\ol{s}}\cdot }, \mand 
P'=\lf P''\rf~.
\]
and we associate with $P$ a set of traces  $\w{tr}(P)$ as in 
section \ref{trace-semantics}.
A general argument shows that
labelled bisimulation is a refinement of trace equivalence.

\begin{proposition}\label{tr-lbis1}
If $P\lbis Q$ then $\w{tr}(P)=\w{tr}(Q)$.
\end{proposition}
\Proof 
We observe that if $P\lbis Q$ and  
$P\act{I/O}P'$ then $Q\act{I/O}Q'$
and $P'\lbis Q'$. From this one can
show that every trace in $\w{tr}(P)$
is in $\w{tr}(Q)$ and conversely.

We recall that
$P\act{I/O} P'$ means 
$P\mid P_I\wact{\tau} P''$, with $P_I = \Pi_{s\in I} \ol{s}$, 
$P''\susp$, $O=\set{s\mid P''\act{\ol{s}}\cdot }$, and
$P'=\lf P''\rf$.
First, note that $P\lbis Q$ implies $P\mid P_I \lbis Q\mid P_I$.
If $(P\mid P_I) \wact{\tau} P''$ and $P''\susp$ then
by $(B1)$ $Q\mid P_I\wact{\tau} Q_1$ and 
$P''\lbis Q_1$.
Moreover, by $(B2)$, $Q_1\wact{\tau} Q''$, $Q''\susp$, 
$P''\lbis Q''$, and $P'= \lf P''\rf \lbis \lf Q''\rf = Q'$. 
By $(B3)$, if $P''\act{\ol{s}}\cdot$ then
$Q''\act{\ol{s}}\cdot$, and conversely. Thus $Q\act{I/O}Q'$. 
\qed  \\

Next, we recast the strong confluence result mentioned in section 
\ref{basic-properties-sec} in the following terms.

\begin{proposition}\label{det-case}
If $P\act{\alpha_{1}}P_1$ and $P\act{\alpha_{2}}P_{2}$ then
either $P_1=P_2$ or $\xst{P_{12}}{(P_1\act{\alpha_{2}}P_{12} \mand
P_2\act{\alpha_{1}}P_{12})}$.
\end{proposition}

We now look at some additional properties that can be derived from
the strong confluence proposition \ref{det-case}.

\begin{lemma}\label{lts-prop2}
\Defitemf{(1)} 
If $P\act{\tau} P_1$, $P\act{s} P_2$, and $\neg P\act{\ol{s}}\cdot$ then
$\xst{P_{12}}{P_1 \act{s} P_{12} \mand P_2\act{\tau} P_{12}}$.

\Defitem{(2)}
If $P\act{s}P'$ and $P\act{\ol{s}}\cdot$ then $P\act{\tau} P'$.

\Defitem{(3)} If $P\act{\tau}P_1$, $P\act{\tau} P_2$ and $P_1\susp$
then $P_1=P_2$.

\Defitemf{(4)}
If $P\wact{\tau} P_1$, $P\wact{\tau} P_2$, $P_1\susp$,
and $P_2\susp$ then $P_1=P_2$.

\Defitem{(5)}
If $P\act{I/O_{1}} P_1$ and $P\act{I/O_{2}} P_2$ then 
$P_1=P_2$ and $O_1=O_2$.
\end{lemma}
\Proof We just check (5).
By (4), if $P\mid P_I \wact{\tau} P'_1$, 
$P'_1\susp$, $P\mid P_I\wact{\tau} P'_2$,
and $P'_2\susp$ then $P'_1=P'_2$. This forces
$P_1 =\lf P'_1\rf = \lf P'_2 \rf = P_2$ and $O_1=O_2$. \qed \\

The following proposition states an interesting consequence of
 confluence.\footnote{One can conceive
non-deterministic extensions of the language where the proposition
fails.}

\begin{proposition}\label{convergence-prop}
$P\lsusp$ if and only if $P\wsusp$.
\end{proposition}
\Proof
By definition, $P\wsusp$ implies $P\lsusp$.
To show the other direction, suppose $P\lsusp$ and
let  $P\act{\alpha_{1}} P_1 \cdots \act{\alpha_{n}} P_n$ be a sequence
of transitions of minimal length leading to a program $P_n$ such
that $P_n\susp$. We build a sequence of internal transitions  $\tau$
leading to a suspended program.
First, we notice that the actions $\alpha_i$ cannot be emission
actions, otherwise a shorter sequence can be found.
Second, we can assume that the last action $\alpha_n$ is an internal
transition $\tau$. Otherwise, if $\alpha_n=s$ 
then either $P_{n-1} \act{\ol{s}}\cdot$ and
then $P_{n-1}\act{\tau}P_n$ by lemma \ref{lts-prop2}(1) or
$\neg P_{n-1} \act{\ol{s}}\cdot$ and then $P_{n-1}\susp$ contradicting
the minimal length hypothesis.

Let us now look at a sequence of transitions:
\begin{equation}\label{trans1}
P \act{s} P_1 \act{\tau} \cdots \act{\tau} P_n \qquad
n\geq 2~.
\end{equation}
where $\neg P \act{\ol{s}}\cdot$ and $\neg P \susp$.
Then we must have $P\act{\tau} P'$ and by lemma \ref{lts-prop2}(1)
there is a $P'_1$ such that $P'\act{s} P'_1$ and $P_1\act{\tau} P'_1$.
By the confluence properties and lemma \ref{lts-prop2}(3),
$P'_1 \wact{\tau} P_n$ in $n-2$ transitions $\tau$.
Thus we have the following sequence of transitions:
\begin{equation}\label{trans2}
P \act{\tau} P'\act{s} P'_1 \wact{\tau} P_n
\end{equation}
The number of $\tau$ transitions that follow the $s$ transition 
is $n-1$ in (\ref{trans1}) and $n-2$ in (\ref{trans2}).
By iterating this reasoning, the input transition $s$ is eventually
removed.
Moreover, the argument is extended to a sequence of transitions
containing several input actions by simply removing the input actions
one after the other proceeding backwards. \qed \\

In view of proposition \ref{convergence-prop}, 
the hypothesis $P\lsusp$ can be replaced by the hypothesis
$P\wsusp$ in condition $(B3)$. Now consider an alternative
definition where the hypothesis $P\lsusp$ is replaced by
the hypothesis $P\susp$. We refer to this condition as
$(B3)^\susp$, call the resulting notion of bisimulation
{\em $\susp$-labelled bisimulation}, and denote  with $\lbissusp$ the
related largest bisimulation. 

\begin{proposition}\label{ddagger-bis}
$\lbis = \lbissusp$.
\end{proposition}

This is a direct consequence of the following lemma
whose proof is delayed to appendix \ref{proof-tau-lemma}.

\begin{lemma}\label{tau-lemma}
\Defitemf{(1)}
If $P\lbis Q$ then $P\lbissusp Q$.

\Defitem{(2)} The relation $\lbissusp$ is reflexive and transitive.

\Defitem{(3)} If $P\act{\tau} Q$ then $P\lbis Q$, $P\lbissusp Q$, and
$\w{tr}(P)=\w{tr}(Q)$.

\Defitem{(4)} $\lbissusp$ is a labelled bisimulation.
\end{lemma}

We rely on this characterisation to show that
bisimulation and trace equivalence collapse; an expected
property of deterministic systems. 
To this end, we note the following properties
of trace equivalence whose proof is given in appendix \ref{proof-tr-lbis2}

\begin{lemma}\label{tr-lbis2}
\Defitemf{(1)}
If $\w{tr}(P)=\w{tr}(Q)$ then $\w{tr}(P\mid (\s{emit}\ s))=
\w{tr}(Q\mid (\s{emit} \ s))$.

\Defitem{(2)} $\cl{R}=\set{(P,Q) \mid \w{tr}(P)=\w{tr}(Q)}$ is a labelled
 bisimulation.
\end{lemma}

From proposition \ref{tr-lbis1} and lemma \ref{tr-lbis2}(2),
we derive the collapse of trace and bisimulation equivalence.

\begin{theorem}\label{tr-bis-equiv}
$P \lbis Q$ if and only if $\w{tr}(P)=\w{tr}(Q)$.
\end{theorem}

\section{Conclusion}
Motivated by recent developments in reactive programming, we have
introduced a revised definition of the SL model  including thread
spawning and recursive definitions.  The revised model is still
confluent and therefore deterministic. We have proposed a simple
static analysis that entails reactivity in the presence of 
recursive definitions and characterised the computational power
of the model with and without signal generation.
Moreover, we have identified a tail recursive core language 
which is built around the \s{present} operator and whose
justification comes directly from the basic design principle of the
SL model. The simplification of the model has been instrumental
to the development of a compositional notion of program equivalence. In further
investigations, we plan to extend this approach to a 
Synchronous Language including data values and name mobility.

\subsection*{Acknowledgements}
The author is indebted to G.~Boudol, F.~Boussinot, I.~Castellani,
and F.~Dabrowski for a number of discussions on the topic of this
paper and for suggesting improvements in its presentation.

\newpage

\appendix

\section{Proofs}\label{proof-appendix}

\subsection{Proof of proposition \ref{prop-unique-decomp}}\label{proof-prop-unique-decomp}
By induction on the structure of $T$ assuming $`;'$ associates
to the right.
If $T=0$ then clearly no decomposition is possible.
If $T\neq 0$ is a redex then take $C=[~]$ and
observe that no other context is possible.
If $T$ has the shape $\Delta;T'$ then take
$C=[~];T'$.
If $T$ has the shape $(\s{watch} \ s \ T')$ and 
$T'\neq 0$ then by inductive hypothesis we
have a unique decomposition $T'=C'[\Delta']$ and
the only possible decomposition for $T$ is obtained
by taking $C=(\s{watch} \ s \ C')$ and $\Delta = \Delta'$.
Finally, if $T= (\s{watch} \ s \ T');T''$ and $T'\neq 0$ then
by inductive hypothesis we
have a unique decomposition $T'=C'[\Delta']$ and 
the only possible decomposition for $T$ is obtained
by taking $C=(\s{watch} \ s \ C');T''$ and $\Delta = \Delta'$.
\qed 

\subsection{Proof of theorem \ref{deterministic-prop}}\label{proof-deterministic-prop}
First we notice that the notion of reduction, suspension, and evaluation
at the end of an instant can be defined up to renaming.

\begin{proposition}\label{red-up-to-renaming}
Suppose $(P_1,E_1) =_\alpha (P_2,E_2)$. 
Then the following holds.

\Defitem{(1)}  If $(P_1,E_1) \act{P''_{1}} (P'_1,E'_1)$ 
then  $(P_2,E_2) \act{P''_{2}} (P'_2,E'_2)$ and 
$(P'_1\union P''_1,E'_1) =_\alpha (P'_2\union P''_2,E'_2)$.

\Defitem{(2)} $(P_1,E_1)\susp$ if and
only if $(P_2,E_2)\susp$.

\Defitem{(3)} If $(P_1,E_1)\susp$ then
$\lf P_1 \rf_{E_{1}} =_\alpha \lf P_2 \rf_{E_{2}}$.

\end{proposition}
\Proof
\Proofitemf{(1)} By case analysis on the reduction.

\Proofitem{(2)} 
Suppose $T_i=C_i[\s{await}\ s_i]$ for $i=1,2$ and $\sigma$ is
a renaming such that $\sigma T_1=T_2$ and $E_1=E_2\comp \sigma$.
Then check that $(T_1,E_1)\susp$ if and only if 
$(T_2,E_2)\susp$.

\Proofitem{(3)} Suppose $(T_1,E_1)=_\alpha (T_2,E_2)$ and 
$(T_1,E_1)\susp$. Proceed by induction on the structure
of $T_1$. \qed \\

Then we check the strong confluence lemma from which
determinism follows.

\begin{lemma}[strong confluence]\label{strong-confluence}
If $(P,E) \act{P''_{1}} (P'_1,E'_1)$,
$(P,E) \act{P''_{2}} (P'_2,E'_2)$,
and $(P'_1\union P''_{1},E'_1)\not=_\alpha (P'_2\union P''_{2},E'_2)$
then there exist $\ol{P}''_{1}, \ol{P}''_{2}, 
P'_{12},E_{12},P'_{21},E_{21}$ such that 
$(P'_1,E'_1) \act{\ol{P}''_{2}} (P'_{12},E_{12})$,
$(P'_2,E'_2) \act{\ol{P}''_{1}} (P'_{21},E_{21})$, and
$(P'_{12}\union P''_1 \union \ol{P}''_2,E_{12}) =_\alpha 
 (P'_{21}\union P''_2 \union \ol{P}''_1,E_{21})$.
\end{lemma}
\Proof
It is convenient to work with a pair $(P,E)$ such that
all bound names are distinct and not in $\w{dom}(E)$.
It is then possible to close the diagram directly taking
$\ol{P}''_{2}=P''_{2}, \ol{P}''_{1}=P''_{1}$,
$P_{12}=P_{21}$, $E_{12}=E_{21}= E_{1}\join E_{2}$,
where:
\[
(E_1 \join E_2)(s) = 
\left\{
\begin{array}{ll}
\true  &\mbox{if }E_1(s)=\true \mbox{ or }E_2(s)=\true \\
\false  &\mbox{otherwise, if }E_1(s)=\false \mbox{ or }E_2(s)=\false \\
\ucl   &\mbox{otherwise.}
\end{array}\right.
\]
We can then derive the initial statement by repeated application of 
proposition \ref{red-up-to-renaming}. \qed

\subsection{Proof of theorem \ref{cps-thm}}\label{proof-cps-thm}
First, it is useful to note the following commutation of substitution
and CPS translation.

\begin{lemma}\label{sub-cps-lemma}
$[\vc{s}/\vc{x}]\dl T \dr(t,\tau) = 
\dl [\vc{s}/\vc{x}]T\dr(t,\tau)$,
assuming $\set{\vc{x}}\inter \w{sig}(t,\tau)=\emptyset$. 
\end{lemma}

\begin{lemma}\label{cps-lemma1}
Suppose $T\rel{R} t$, and $(T,E)\act{P} (T', E')$. Then 
$T=C[\Delta]$ for some context $C$ and redex $\Delta$ 
and exactly one of the following cases arises.

\Defitem{(1)} $\Delta::=0;T'' \Alt (\s{watch} \ s \ 0)$.
Then $P=\emptyset$, $E=E'$, and
$t=\dl T \dr (0,\epsilon) = \dl T'\dr(0,\epsilon)$.

\Defitem{(2)} $\Delta::=\s{thread} \ T''$.
Then $P= \pset{T''}$, $E=E'$, and
$(t,E)=(\dl T \dr(0,\epsilon),E)
\act{\pset{\dl T'' \dr(0,\epsilon)}} (\dl T'\dr(0,\epsilon),E)$.

\Defitem{(3)} $\Delta ::= \s{emit} \ s \Alt \nu s \ T'' \Alt A(\vc{s})$.
Then $P=\emptyset$ and 
$(t,E) = (\dl T \dr(0,\epsilon),E)  \act{\emptyset} (\dl
T'\dr(0,\epsilon),E')$.

\Defitem{(4)} $\Delta::= \s{await} \ s$ and $t= \dl T \dr(0,\epsilon)$.
Then $P=\emptyset$, $E=E'$, and 
$(t,E)  \act{\emptyset} (\dl T'\dr(0,\epsilon),E)$.

\Defitem{(5)} $\Delta::=\s{await} \ s$ and $t=A$ where
$A= \dl T \dr(0,\epsilon)$. Then $P=\emptyset$, $E=E'$, and 
$(t,E) (\act{\emptyset})\cdot (\act{\emptyset}) (\dl T'\dr(0,\epsilon),E)$.
\end{lemma}

\Proof We denote with $\pi_1,\pi_2$ the first and
second projection, respectively.

\Proofitem{(1)}
If $\Delta = 0;T$ then 
\[
\begin{array}{ll}

\dl C[0;T] \dr(0,\epsilon)  \\

= \dl 0; T \dr (\dl C \dr (0,\epsilon))
&(\mbox{by proposition \ref{cxt-cps}}) \\

= \dl T \dr (\dl C \dr (0,\epsilon))
&(\mbox{by CPS definition}) \\

= \dl C[T] \dr (0,\epsilon) 
&(\mbox{by proposition \ref{cxt-cps}})~.
\end{array}
\]
If $\Delta = \s{watch} \ s  \ 0$ 
let $(t,\tau) = \dl C \dr (0,\epsilon)$.
Then 
\[
\begin{array}{ll}

\dl C[\s{watch} \ s  \ 0] \dr(0,\epsilon)  \\

= \dl \s{watch} \ s  \ 0 \dr (t,\tau)
&(\mbox{by proposition \ref{cxt-cps}}) \\

= \dl 0 \dr (t, \tau\cdot (s, t))
&(\mbox{by CPS definition}) \\

= t  
&(\mbox{by CPS definition}) \\

= \dl 0 \dr (t,\tau) 
&(\mbox{by CPS definition}) \\

= \dl C[0] \dr (0,\epsilon)
&(\mbox{by proposition \ref{cxt-cps}})~.
\end{array}
\]
\Proofitemf{(2)}
We observe: 
\[
\begin{array}{ll}

\dl C[\s{thread} \ T''] \dr(0,\epsilon)  \\

= \dl \s{thread} \ T'' \dr (\dl C \dr (0,\epsilon))
&(\mbox{by proposition \ref{cxt-cps}}) \\

= \s{thread} \  \dl T'' \dr(0,\epsilon).
\pi_1 (\dl C \dr (0,\epsilon))
&(\mbox{by CPS definition}) \\

= \s{thread} \  \dl T'' \dr(0,\epsilon).
\dl 0 \dr (\dl C \dr (0,\epsilon))
&(\mbox{by CPS definition}) \\

\act{\pset{\dl T'' \dr(0,\epsilon)}} \dl C[0] \dr(0,\epsilon)
&(\mbox{by }(t_5)\mbox{ and proposition }\ref{cxt-cps})
\end{array}
\]
\Proofitemf{(3)}
The cases where $\Delta=(\s{emit} \ s)$ or 
$\Delta =  (\nu  s \ T)$ are straightforward. 
Suppose $\Delta=A(\vc{s})$. Assume 
$(t,\tau) = \dl C \dr (0,\epsilon)$, $\w{sig}(t,\tau) = \set{\vc{s'}}$
and $A(\vc{x}) = T$ with $\set{\vc{x}}\inter \set{\vc{s'}} = \emptyset$.
We consider the equation 
$A^{(t,\tau)}(\vc{x}) = \dl T \dr (t,\tau)$ where 
we rely on the convention that the parameters $\vc{s'}$ are 
omitted. Now we have:
\[
\begin{array}{ll}
\dl C[A(\vc{s})] \dr (0,\epsilon)  \\

= \dl A(\vc{s}) \dr (\dl C \dr (0,\epsilon)) 
&(\mbox{by proposition \ref{cxt-cps}}) \\

= A^{(t,\tau)}(\vc{s}) 
&(\mbox{by CPS definition})\\

\act{\emptyset}
[\vc{s}/\vc{x},\vc{s'}/\vc{s'}]\dl T \dr (t,\tau) \\

= \dl [\vc{s}/\vc{x}]T \dr (t,\tau) 
&(\mbox{by substitution lemma }\ref{sub-cps-lemma})\\

= \dl [\vc{s}/\vc{x}]T \dr (\dl C \dr (0,\epsilon)) \\

= \dl C[[\vc{s}/\vc{x}]T] \dr (0,\epsilon)
&(\mbox{by proposition \ref{cxt-cps}}).

\end{array}
\]
\Proofitemf{(4)}
We observe:
\[
\dl C[\s{await} \ s ] \dr(0,\epsilon) = 
\dl \s{await}\ s \dr (\dl C \dr (0,\epsilon)) =
\s{present} \ s \  t \ b
\]
where $t=\pi_{1}(\dl C \dr (0,\epsilon)) =
\dl C[0]\dr(0,\epsilon)$ and
$(\s{present} \ s \ t \ b,E) \act{\emptyset} (t,E)$.

\Proofitem{(5)} First unfold $A(\vc{s})$ and then proceed
as in case (4). \qed \\

Thus if $T\rel{R} t$ and $T$ reduces then $t$ can match the reduction
and stay in the relation. The proofs of the following three lemma
\ref{cps-lemma2}, \ref{lemma-cps3}, and \ref{lemma-cps4} rely on
similar arguments.  First, we analyse the situation where $t$ reduces.

\begin{lemma}\label{cps-lemma2}
Suppose $T\rel{R} t$, and $(t,E)\act{p} (t', E')$. Then
$T=C[\Delta]$ and  exactly one of the following cases arises.

\Defitem{(1)} $\Delta::=\s{await} \ s$ and $t=A$ where
$A= \dl T \dr(0,\epsilon)$. Then $p=\emptyset$, $E=E'$ and 
$T\rel{R} t'$.

\Defitem{(2)} $\Delta::= \s{await} \ s$ and $t= \dl T \dr(0,\epsilon)$.
Then $p=\emptyset$, $E=E'$, and 
$(T,E)  \act{\emptyset} (T',E)$ with $t'= \dl T'\dr(0,\epsilon)$.

\Defitem{(3)} $\Delta::=\s{thread} \ T''$.
Then $p= \pset{\dl T''\dr(0,\epsilon)}$, $E=E'$, and
$(T,E) \act{\pset{T''}} (T',E)$ with $t'=\dl T'\dr(0,\epsilon)$.

\Defitem{(4)} $\Delta ::= \s{emit} \ s \Alt \nu s \ T'' \Alt A(\vc{s})$.
Then $p=\emptyset$, $t=\dl T \dr(0,\epsilon)$, and
$(T,E)  \act{\emptyset} (T',E')$ with 
$t'= \dl T'\dr(0,\epsilon)$.

\Defitem{(5)} $\Delta::=0;T'' \Alt (\s{watch} \ s \ 0)$.
Then $p=\emptyset$, $E=E'$, $t=\dl T \dr(0,\epsilon)$
$(T,E) \act{\emptyset} (T',E)$, 
$t= \dl T '\dr (0,\epsilon)$, and $T'$ is smaller than $T$.
\end{lemma}

Thus if $T\rel{R} t$ and $t$ reduces then $T$ can match the reduction and
stay in the relation. In the worst case, the number of reductions $T$
has to make is proportional to its size. This is because 
case (5) shrinks the thread. 

\begin{lemma}\label{lemma-cps3}
If $T\rel{R} t$ and $(T,E)\susp$ then 
exactly one of the following cases arises.

\Defitem{(1)}
$t = \dl T \dr(0,\epsilon)$. Then $(t,E)\susp$.

\Defitem{(2)}
$T=C[\s{await} \ s]$, $t= A$, 
and $A= \dl T \dr(0,\epsilon)$. Then
$(t,E)\act{\emptyset} (\dl T \dr(0,\epsilon),E)$ and 
$(\dl T \dr(0,\epsilon),E)\susp$.

\end{lemma}

Thus if $T\rel{R} t$ and $(T,E)$ is suspended then $(t,E)$ is
suspended too possibly up to an unfolding. 

\begin{lemma}\label{lemma-cps4}
If $T\rel{R} t$ and $(t,E)\susp$ then 
$t=\dl T \dr(0,\epsilon)$ and exactly one of the following cases arises.

\Defitem{(1)}
$T=0$ or $T=C[\s{await} \ s]$ and $(T,E)\susp$.

\Defitem{(2)}
$T=C[\Delta]$, $\Delta::=0;T'' \Alt (\s{watch} \ s \ 0)$.
Then $(T,E) \act{\emptyset} (C[0],E)$ and
$t=\dl C[0]\dr(0,\epsilon)$.
\end{lemma}

Thus if $T\rel{R} t$ and $(t,E)$ is suspended then
$(T,E)$ is suspended too possibly up to the reduction
of redexes $0;T''$ or  $(\s{watch} \ s \ 0)$.
Again the number of these reductions is at most proportional
to the size of $T$. 
Next we look at the computation at the end of the instant.

\begin{lemma}\label{lemma-cps5}
If $T\rel{R} t$, $(T,E)\susp$,  and $(t,E)\susp$ then 
$\lf T \rf_E \rel{R} \lf t \rf_E$.
\end{lemma}
\Proof Exactly one of the following cases arises.

\Proofitem{(1)}
$T=t=0= \lf T \rf_E = \lf t \rf_E$.

\Proofitem{(2)} $T=C[\s{await} \ s]$, $t=\dl T \dr(0,\epsilon)$.
We have to explicit the structure of $t$ and relate it to the
structure of the context.
First, we notice that the context $C$ can be written
in the general form
\[
C=(\s{watch} \ s_1  \cdots (\s{watch} \ s_n \ [~]U_{n+1})U_n \cdots )U_1 
\]
where $U_i::= \epsilon \Alt \  ;T_i$ so that the presence of
$U_i$ is optional.
Then we claim that $t$ can be written as:
\[
t=\s{present} \ s \ t_{n+1} (\s{ite} \ s_1 \ t_1 \ \cdots (\s{ite} \
s_n \ t_n A)\cdots), \quad   A = t 
\]
where $t_i$ is defined inductively as follows:
\[
\begin{array}{lll}
t_0 &= 0, \\
\tau_0 &= \epsilon \\

t_{i+1} &= \left\{ \begin{array}{ll}
          \dl T_{i+1} \dr (t_i,\tau_{i}) &\mbox{if }U_{i+1} = \ ;T_{i+1} \\
          t_i                          &\mbox{otherwise} 
          \end{array} \right.
&\mbox{for } i=0,\ldots,n \\

\tau_{i+1} &= \tau_i \cdot (s_{i+1},t_{i+1}) 
&\mbox{for }i=0,\ldots,n-1 
\end{array}
\]
In particular, we have $\dl C \dr (0,\epsilon) = (t_{n+1},\tau_n)$.
Now two subcases can arise.

\Proofitem{(2.1)} $E(s_1)=\cdots = E(s_n) = \false$.
Then $\lf T \rf_E = T$ and $\lf t \rf_E = A$ so
that thanks to the second clause in the definition of $\cl{R}$ 
we have $\lf T \rf_E \cl{R} \lf t \rf_E$.

\Proofitem{(2.2)} $E(s_1)=\cdots = E(s_{i-1}) = \false$ and
$E(s_{i})=\true$. Then 
\[
\lf T \rf_E = 
(\s{watch} \ s_1  \cdots (\s{watch} \ s_{i-1} \ 0 \ U_{i})U_{i-1} \cdots )U_1,
\ \mand \ \dl \lf T \rf_E \dr(0,\epsilon) = t_i = \lf t \rf_E~. \qquad \Box
\]

To summarise, we have shown that the relation $\cl{R}$ acts
as a kind of weak bisimulation with respect to reduction and
suspension and that it is preserved by the computation at the
end of the instant.
Note that the relation $\cl{R}$ is immediately extended to programs in the
source and target language by saying that the source program $P$ is
related to the target program $p$ if there is a bijection $i$ between
the threads in $P$ and those in $p$ such that if $i(T)=t$ then
$T\rel{R} t$.

\begin{lemma}\label{lemma-cps6}
Suppose $P\rel{R} p$. Then for every environment $E$:

\Defitem{(1)}
If $(P,E) (\arrow)^* (P',E')$ and $(P',E')\susp$ then
for some $p'$
$(p,E) (\arrow)^* (p',E')$, $(p',E')\susp$, and
$\lf P' \rf_{E'}\rel{R} \lf p'\rf_{E'}$.

\Defitem{(2)} Vice versa,
if $(p,E) (\arrow)^* (p',E')$ and $(p',E')\susp$ then
for some $P'$
$(P,E) (\arrow)^* (P',E')$, $(p',E')\susp$, and
$\lf P' \rf_{E'}\rel{R}\lf p'\rf_{E'}$.

\end{lemma}

From lemma \ref{lemma-cps6} we derive that if $P\rel{R} p$ then
$\w{tr}(P) = \w{tr}(p)$ and in particular that $\w{tr}(P) = \w{tr}(\dl
P \dr)$ as required.

\subsection{Proof of proposition \ref{bounded-cxt-prop}}\label{proof-bounded-cxt-prop}
Let $X$ be a finite set of thread identifiers.
We define its {\em depth} as the length of the longest
descending chain with respect to $\succ$.
Consider an equation.
$A(\vc{x})=T$. The function $\w{Call}(T,\epsilon)$
implicitly associates a label $\ell\in \set{\epsilon,\kappa}$
with every occurrence of a thread identifier in $T$.
Next consider a related equation 
$A^{(t,\tau)}(\vc{x}) = \dl T \dr(t,\tau)$ and an occurrence
of a thread identifier $B$ in $T$. Two situations may arise:
(1) The label associated with the occurrence of $B$ is $\kappa$
and then $A\succ B$.
(2) The label associated with the occurrence of $B$ is $\epsilon$
and then $A\succeq B$ and moreover the index $(t',\tau')$ of $B$ 
in the CPS translation is either $(0,\epsilon)$ or $(t,\tau)$.

Then to compute the system of recursive equations associated
with the CPS translation proceed as follows.
First, compute the equations of `index' $(0,\epsilon)$,
{\em i.e.}, those of the shape
$A^{(0,\epsilon)}(\vc{x}) = \dl T \dr(0,\epsilon)$
and collect all the thread identifiers $A^{(t,\tau)}$ 
occurring on the right hand side with an index $(t,\tau)$ 
different from $(0,\epsilon)$.
Continue,  by computing the equations
$A^{(t,\tau)} = \dl T \dr(t,\tau)$ for the new indexes $(t,\tau)$.
Then collect again the identifiers with new indexes.
At each step the depth of the finite set of thread identifiers
with new indexes decreases. Thus this process terminates
with a finite number of recursive equations. \qed

\subsection{Proof of theorem \ref{undecidability-thm}}\label{proof-undecidability-thm}
We start by describing the simulation of simple deterministic {\em push down
automata}.  The empty stack is represented by the symbol $Z$.  The stack
alphabet has only one symbol $S$.  
A configuration of an automaton is a pair 
$(q,S\cdots S Z)$ composed of a state and a stack,
and its possible transitions are:
\[
\begin{array}{ll}

(q,w)\mapsto (q',Sw)  &(\mbox{increment}) \\

(q,Sw)\mapsto (q',w)  &(\mbox{decrement}) \\

(q,w)\mapsto \left\{\begin{array}{ll}
                   (q',w)     &w=Z \\
                   (q'',w)    &w\neq Z
                    \end{array} \right. 
 &(\mbox{test zero})

\end{array}
\]
We introduce as many thread identifiers as states.
Each of these thread identifiers has parameters
$\w{inc}$, $\w{dec}$, $\w{zero}$, $\w{ack}$ which we omit.
Depending on the instructions associated with the state,
we introduce one of the following equations:
\[
\begin{array}{lll}
q  &= (\s{emit} \ \w{inc}); (\s{await} \ \w{ack}); \s{pause}; q'
   &(\mbox{increment}) \\

q  &= (\s{emit} \ \w{dec}); (\s{await} \ \w{ack}); \s{pause}; q' 
   &(\mbox{decrement})\\

q  &= (\s{present} \ \w{zero} \ (\s{pause};q') \ q'')
   &(\mbox{test zero})

\end{array}
\]
Note that the control starts at most one operation per instant
and that it waits for the completion of the operation before
proceeding to the following one. 

Next we represent the stack.  This is similar to what is done, {\em
e.g.}, in {\sc CCS} \cite{Milner89}.  We abbreviate with $\vc{s}$ a vector of $5$ signals
$\w{dec},\w{inc},\w{zero},\w{ack},\w{abort}$.  A thread $Z$ depends on
such a vector for interactions on the `left'.  A thread $S$ (or $S_+,
S_r, S_l$) depends on a pair of vectors $\vc{s}, \vc{s'}$ for
interactions on the `left'and on the `right', respectively.
\[
\begin{array}{ll}
Z(\vc{s}) 
&= (\s{watch} \ \w{abort} \ (\s{emit} \ \w{zero});  \\
   &\qquad (\s{present} \ \w{inc} \\
   &\qquad\quad (\s{emit} \ \w{ack}); \s{pause};
   (\nu  \vc{s'} \ (\s{thread} \ S(\vc{s},\vc{s'}), Z(\vc{s'})))\\
   &\qquad\quad (\s{thread} \ Z(\vc{s})))) \\ \\

S(\vc{s},\vc{s'}) &= (\s{thread} \ \\
&\quad (\s{watch} \ \w{dec} \ (\s{await} \ \w{inc}); \s{pause};
       (\s{thread} \ S_+(\vc{s},\vc{s'}))), \\
&\quad (\s{watch} \ \w{inc} \ (\s{await} \ \w{dec}); \s{pause};
       (\s{thread} \ S_r(\vc{s},\vc{s'}))) ) \\ \\

S_+(\vc{s},\vc{s'}) &= (\nu  \vc{s''} \ (\s{emit} \ \w{ack}); \ 
        (\s{thread} \ S(\vc{s},\vc{s''}), S(\vc{s''},\vc{s'}))) \\ \\

S_r(\vc{s},\vc{s'}) &= (\s{present} \ \w{zero'} \ 
(\s{emit} \ \w{abort'}); \s{pause}; (\s{emit} \ \w{ack}); Z(\vc{s}) \\
&\quad (\s{emit}\ \w{dec'});  S_l(\vc{s},\vc{s'}) \\ \\

S_l(\vc{s},\vc{s'}) &= (\s{await} \ \w{ack'}); \s{pause}; 
(\s{emit} \ \w{ack}); S(\vc{s},\vc{s'})

\end{array}
\]
A configuration $(q,S\cdots S Z)$ of the automaton is mapped to a
program which is essentially equivalent to:
$(\nu  \vc{s}_0,\ldots,\vc{s}_n \ 
(\s{thread}\ q(\vc{s}_0),S(\vc{s}_{0},\vc{s}_{1}),\ldots,
             S(\vc{s}_{n-1},\vc{s}_{n}),Z(\vc{s}_{n})))$.
It is not difficult to check that the program can simulate the
transitions of the automata (and this is all we need to check since
the program is deterministic!).  The more complex dynamics, is
introduced by the decrement.  Roughly, the decrement of a stack
represented by the threads $S, S,S,Z$ goes through the following
transformations:
\[
\begin{array}{llll}
S,S,S,Z \arrow S_r,S,S,Z \arrow S_l,S_r,S,Z \arrow S_l,S_l,S_r,Z 
\arrow S_l,S_l,Z \arrow S_l,S,Z \arrow S,S,Z
\end{array}
\]
There is a wave from left to right that transforms $S$ into $S_l$,
when the wave meets $Z$, it aborts $Z$, transforms the rightmost 
$S$ into $Z$, and produces a wave from right to left that turns 
$S_l$ into $S$ again. 
The simulating program can be put in tail recursive form via
the CPS translation. In particular, note that all recursive calls in the scope of
a $\s{watch}$ are under
a $\s{thread}$ statement that has the effect of resetting 
the evaluation context.
Finally, we remark that 
the simulation of deterministic push down automata can be easily
generalised to deterministic two counters machines by simply letting
the control operate on two distinct stacks.   \qed

\subsection{Proof of proposition \ref{lts-prop}}\label{proof-lts-prop}
\Proofitemf{(1)}
By induction on the proof of $P\act{\ol{s}}P'$.


\Proofitem{(2)} If $P\act{\ol{s}}\cdot$ then $P$ has the shape
$D[\s{emit} \ s]$ for a suitable context $D$ built out of restrictions
and parallel compositions. It is easily checked
that after a transition the emission $\s{emit} \ s$ is still
observable.

\Proofitem{(3)} By induction on the proof of 
$P\act{s} P'$. \qed

\subsection{Proof of lemma \ref{lemma-sbis}} \label{proof-lemma-sbis}
Most properties follow by routine verifications.
We just highlight some points.

\Proofitem{(1)}
Recalling that $P\sbis Q$ and $P\susp$ implies $Q\susp$.

\Proofitem{(2)}
Condition $(S1)$ entails conditions $(B1)$, $(B3)$, and $(L2)$,
while condition $(S2)$ (with $(S1)$) entails conditions
$(B2)$ and $(L1)$.

\Proofitem{(3)}
Introduce a notion of normalised program where 
parallel composition associates to the left,
all restrictions are carried at top level, 
and $0$ programs are removed.
Then define a relation $R$ where two programs are
related if their normalised forms are identical
up to bijective permutations of the restricted names
and the parallel components. A pair of programs equated by the laws
under consideration is in $R$.
Show that $R$ is a strong labelled bisimulation. 

\Proofitem{(4)}
Show that 
$\set{(P\mid S, Q\mid S) \mid P\sbis Q}$ is a strong
labelled bisimulation where $S$ is defined 
as in the statement.

\Proofitem{(5)} Direct diagram chasing.

\Proofitem{(6)} 
We reason up to $\sbis$.

\Proofitem{(7)}
We show $\set{(P_1\mid Q, P_2\mid Q) \mid P_1 \sbis P_2}$
is a strong labelled bisimulation up to $\sbis$. 
Let us focus on condition $(S2)$.
Let $X= \set{s'\mid (P_1\mid P_2) \act{\ol{s'}} \cdot }$ and
let $S'$ be the parallel composition of 
the emissions $(\s{emit} \ s)$ where $s\in X$.
Suppose $(P_1\mid Q \mid S) \susp$. Then
we note that
$P_1\mid Q \mid S \sbis (P_1\mid S' \mid S) \mid (Q \mid S'\mid S)$
and 
$\lf P_1\mid Q \mid S \rf \sbis
\lf P_1\mid S' \mid S \rf \mid \lf Q \mid S'\mid S \rf$.
A similar remark applies to $P_2\mid Q$. Then we can conclude
by reasoning up to $\sbis$.  \qed 

\subsection{Proof of lemma \ref{lemma-bis-congr}}\label{proof-lemma-bis-congr}
\Proofitemfb{(1)}
We show that the relation 
$R=\lbis \union 
\set{( \ P\mid (\s{emit} \ s), Q\mid (\s{emit} \ s)\  ) \mid P\lbis  Q}$
is a labelled bisimulation up to $\sbis$. 
We assume $P\lbis Q$ and we analyse the conditions 
$(B1-3)$ and $(L1-2)$.

\Proofitem{(B1)} Suppose
$P\mid (\s{emit} \ s) \act{\tau} P'\mid (\s{emit} \ s)$. If the
action $\tau$ is performed by $P$ then the hypothesis and condition
(B1) allow to conclude. Otherwise, suppose $P\act{s} P'$. Then we
apply the hypothesis and condition $(L2)$. Two cases may arise:
(1) If $Q\wact{s} Q'$ and $P'\lbis Q'$ then the conclusion is
immediate.
(2) If $Q\wact{\tau} Q'$ and $P'\lbis Q'\mid (\s{emit} \ s)$ then
we note that $Q'\mid (\s{emit} \ s) \sbis (Q'\mid (\s{emit} \ s)) \mid 
(\s{emit} \ s)$ and we close the diagram up to $\sbis$.


\Proofitem{(B3)} Suppose 
$P\mid (\s{emit} \ s) \act{\ol{s'}} \cdot$ and 
$P\mid (\s{emit} \ s) \lsusp$.
If $s=s'$ then $Q\mid (\s{emit} \ s) \act{\ol{s'}} \cdot$ and we
are done. Otherwise, it must be that $P\act{\ol{s'}}\cdot$.
Moreover, $P\lsusp$. Then $P\lbis Q$ and condition $(B3)$
imply that $Q\wact{\tau} Q'\act{\ol{s'}}\cdot$, and 
$P\lbis Q'$. 
Hence 
$Q\mid (\s{emit} \ s) \wact{\tau} Q'\mid (\s{emit} \ s) \act{\ol{s'}}\cdot$
and we can conclude.

\Proofitem{(L1)} Suppose $S=(\s{emit} \ s_1) \mid \cdots \mid (\s{emit} \ s_n)$.
Define $S'= (\s{emit} \ s)\mid S$. Then $P\lbis Q$ and condition $(L1)$
applied to $S'$ allows to conclude.

\Proofitem{(L2)} Suppose 
$P\mid (\s{emit} \ s)\act{s'} P'\mid (\s{emit} \ s)$.
Necessarily $P\act{s'} P'$. Given $P\lbis Q$ and condition $(L2)$ 
two cases may arise: 
(1) $Q\wact{s'} Q'$ and $P'\lbis Q'$. Then the conclusion is
immediate.
(2) $Q\wact{\tau} Q'$ and $P'\lbis Q'\mid  (\s{emit} \ s')$.
Then $Q\mid (\s{emit} \ s) \wact{\tau} Q'\mid (\s{emit} \ s)$ 
and we observe that 
$(Q'\mid (\s{emit} \ s)) \mid (\s{emit} \ s') \sbis
(Q'\mid (\s{emit} \ s')) \mid (\s{emit} \ s)$ thus closing
the diagram up to $\sbis$.

\Proofitemb{(2)} 
It is easily checked that the identity relation is a labelled
bisimulation. Reflexivity follows. 
As for transitivity, we check that the relation 
$\lbis \comp \lbis$ is a labelled bisimulation up to $\sbis$.

\Proofitem{(B1-3,L1)}  These cases are direct. For $(B3)$,
recall proposition \ref{l-suspension}(2).

\Proofitem{(L2)} Suppose $P_1 \lbis P_2 \lbis P_3$ and 
$P_1 \act{s} P'_1$. Two interesting cases arise when either
$P_2$ or $P_3$ match an input action with an internal transition.
(1) Suppose first $P_2 \wact{\tau} P'_2$ and 
$P_1 \lbis P'_2 \mid (\s{emit} \ s)$. By $P_2\lbis P_3$ and
repeated application of $(B1)$ we derive that
$P_3 \wact{\tau} P'_3$ and $P'_2 \lbis P'_3$. 
By property (1) the latter implies that
$P'_2 \mid (\s{emit} \ s) \lbis P'_3 \mid (\s{emit} \ s)$ and
we combine with $P_1 \lbis P'_2 \mid (\s{emit} \ s)$ to conclude.
(2) Next suppose 
$P_2 \wact{\tau} P^{1}_{2} \act{s} P^{2}_{2} \wact{\tau} P'_2$ and
$P_1 \lbis P'_2$. Suppose that $P_3$ matches these transitions as
follows:
$P_3 \wact{\tau} P^{1}_{3} \wact{\tau} P^{2}_{3}$, 
$P^{2}_{2}\lbis P^{2}_{3} \mid (\s{emit} \ s)$, and moreover
$P^{2}_{3} \mid (\s{emit} \ s) \wact{\tau} P'_3  \mid (\s{emit} \ s)$ with
$P'_2 \lbis P'_3\mid (\s{emit} \ s)$. Two subcases may arise:
(i) $P^{2}_{3} \wact{\tau}  P'_3$. Then we have 
$P_3 \wact{\tau} P'_3$, $P'_2 \lbis  P'_3\mid (\s{emit} \ s)$ and
we can conclude.
(ii) $P^{2}_{3} \wact{s}  P'_3$. Then
 we have $P_3 \wact{s} P'_3$ and 
$P'_2 \lbis P'_3 \mid (\s{emit} \ s) \sbis P'_3$.

\Proofitemb{(3)}
We show that $\set{(\nu s \ P, \nu s \ Q) \mid P\lbis Q}$ is a 
labelled bisimulation up to $\sbis$.

\Proofitem{(B1)}
If $\nu s \ P\act{\tau} P''$ then $P''= \nu s P'$ and
$P\act{\tau} P'$. From $P\lbis Q$  and $(B1)$ we derive
$Q\wact{\tau} Q'$ and $P'\lbis Q'$. Then
$\nu s\ Q \wact{\tau} \nu s \ Q'$ and we conclude.


\Proofitem{(B3)}
If $\nu s \ P \act{\ol{s'}} \cdot$ ($s\neq s'$) then 
$P\act{\ol{s'}}\cdot$. From $P\lbis Q$  and $(B3)$ we derive
$Q\wact{\tau} Q'$, $Q'\act{\ol{s'}}\cdot$, and 
$P\lbis Q'$. To conclude, note that $\nu s \ Q\wact{\tau} \nu s \ Q'$ and
$\nu s \ Q'\act{\ol{s'}}\cdot$.

\Proofitem{(L1)}
Let $S=(\s{emit} \ s_1) \mid \cdots \mid (\s{emit} \ s_n)$ with $s\neq s_i$
for $i=1,\ldots,n$. If $((\nu s \ P) \mid S) \susp$ then 
$(P \mid S) \susp$. From $P\lbis Q$ and $(L1)$ we derive
$(Q\mid S) \wact{\tau} (Q'\mid S)$, $(Q'\mid S) \susp$, 
$(P \mid S) \lbis (Q'\mid S)$, and 
$\lf P \mid S\rf \lbis \lf Q'\mid S \rf$.
This implies that $((\nu s \ Q) \mid S) \wact{\tau} ((\nu s \ Q')\mid S)$
and $((\nu s \ Q')\mid S) \susp$. We observe that 
$((\nu s \ P) \mid S) \sbis \nu s \ (P\mid S)$,
$((\nu s \ Q') \mid S) \sbis \nu s \ (Q'\mid S)$,
$\lf (\nu s \ P) \mid S \rf \sbis \nu s \ \lf P \mid S \rf$, and
$\lf (\nu s \ Q') \mid S \rf \sbis \nu s \ \lf Q' \mid S \rf$.
Then we can close the diagram up to $\sbis$.

\Proofitem{(L2)}
Suppose $\nu s \ P \act{s'} P''$. Then $s\neq s'$ and
$P''= \nu s \ P'$ with $P\act{s'} P'$.
From $P\lbis Q$ and $(L2)$ two cases may arise. 
(1) If $Q\wact{s'} Q'$ and $P'\lbis Q'$ then
$\nu s \ Q \wact{s'} \nu s \ Q'$ and we are done.
(2) If $Q\wact{\tau} Q'$ and $P'\lbis Q'\mid (\s{emit} \ s')$ then
$\nu s \ Q \wact{\tau} \nu s\ Q'$ and we note that 
$\nu s \ Q' \mid (\s{emit} \ s') \sbis \nu s \ (Q' \mid (\s{emit} \ s'))$
thus  closing the diagram up to $\sbis$.

\Proofitemb{(4)}
We show that $R = \set{(P_1\mid Q, P_2 \mid Q) \mid P_1 \lbis P_2} 
\union \lbis$
is a labelled bisimulation up to $\sbis$.

\Proofitem{(B1)}
Suppose $(P_1\mid Q) \act{\tau} P'$.

\Defitem{(B1)[1]}
If the $\tau$ transition is due to $P_1$ or $Q$ then 
the corresponding $P_2$ or $Q$ matches the transition and we are done.

\Defitem{(B1)[2]} Otherwise, suppose  $P_1\act{s} P'_{1}$ and
$Q \act{\ol{s}} Q$. 

\Defitem{(B1)[2.1]} If $P_2 \wact{s} P'_2$ and $P'_1\lbis P'_2$ then
$(P_2\mid Q) \wact{\tau} (P'_2 \mid Q)$ and we are done.

\Defitem{(B1)[2.2]} If $P_2\wact{\tau} P'_2$ and $P'_1\lbis (P'_2\mid
(\s{emit} \ s))$ then 
$(P_2\mid Q) \wact{\tau} (P'_2\mid Q)$ and 
$((P'_2\mid Q)\mid (\s{emit} \ s)) \sbis ((P'_2 \mid (\s{emit} \
s))\mid Q)$ so that we close the diagram up to $\sbis$.

\Defitem{(B1)[3]} Otherwise, suppose $P_1\act{\ol{s}} P_1$ and 
$Q\act{s} Q'$. 

\Defitem{(B1)[3.1]}
If $\neg P_1 \lsusp$ then by lemma \ref{l-suspension},
$\neg (P_1\mid Q)\lsusp$,
$\neg (P_1\mid Q') \lsusp$, 
$\neg P_2 \lsusp$, $\neg (P_2\mid Q) \lsusp$.
Therefore $(P_1\mid Q') \lbis (P_2 \mid Q)$.

\Defitem{(B1)[3.2]} If $P_1\lsusp$ then $P_2 \wact{\ol{s}}P'_2$ 
and $P_1 \lbis P'_2$. Hence
$(P_2 \mid Q) \wact{\tau} (P'_2 \mid Q')$ and 
$(P_1 \mid Q') \rel{R} (P'_2 \mid Q')$.


\Proofitem{(B3)} Suppose $(P_1\mid Q)\lsusp$.

\Defitem{(B3)[1]}
Suppose $P_1\act{\ol{s}}\cdot$.
Then $P_1\lsusp$ and by $(B3)$ $P_2 \wact{\tau} P'_2
\act{\ol{s}}\cdot$ and $P_1\lbis P'_2$.
Thus $(P_2 \mid Q) \wact{\tau} (P'_2\mid Q) \act{\ol{s}} \cdot$
and we can conclude.

\Defitem{(B3)[2]}
Suppose $Q\act{\ol{s}}$.
Then $(P_2\mid Q)\act{\ol{s}}$ and we are done.

\Proofitem{(L1)} 
Suppose $(P_1\mid Q \mid S)\susp$.
Then $(P_1\mid S)\susp$ and from
$P_1\lbis P_2$ we derive 
$(P_2 \mid S) \wact{\tau} (P'_{2} \mid S) \susp$ and
$(P_1 \mid S) \lbis (P'_{2} \mid S)$.
In particular, 
$\set{s \mid P_1 \mid S \act{\ol{s}}\cdot} = 
\set{s \mid P'_2 \mid S \act{\ol{s}}\cdot}$.
We can also derive that 
$(P_2\mid Q \mid S) \wact{\tau} (P'_2 \mid Q \mid S)$,
however $(P'_2 \mid Q \mid S)\susp$ may fail because of 
a synchronisation of $P'_2$ and $Q$ on some signal which
is not already in $S$. 
Then we consider $S'$ as the parallel composition of
emissions $(\s{emit} \ s)$ where $(P_1\mid Q) \act{\ol{s}}\cdot$.
By lemma \ref{lemma-sbis}, we derive that:
\[
\begin{array}{lll}
\w{(i)} &(P_1\mid Q \mid S) &\sbis (P_1\mid S \mid S') \mid (Q \mid S \mid S')\quad \mand \\
\w{(ii)} &(P'_2\mid Q \mid S) &\sbis (P'_2\mid S \mid S') \mid (Q \mid S \mid S')~.
\end{array}
\]
We also observe that $(P_1 \mid S \mid S') \susp$. 
Together with $(P_1 \mid S) \lbis (P'_2 \mid S)$ this implies by $(L1)$ 
$(P'_2 \mid S \mid S')\wact{\tau} (P''_2 \mid S \mid S')\susp$,
$(P_1 \mid S \mid S')\lbis (P''_2 \mid S \mid S')$, and
$\lf P_1 \mid S \mid S' \rf \lbis \lf P''_2 \mid S \mid S'\rf$.
Now it must be that 
$((P''_2 \mid S \mid S') \mid (Q \mid S \mid S'))\susp$ 
because the left component already emits all the signals that
could be emitted by the right one (and vice versa).
By conditions $(S1-2)$ and \w{(ii)} we have that
$(P'_2 \mid Q \mid S) \wact{\tau} (P'''_2 \mid Q \mid S) \susp$
and $(P'''_2 \mid Q \mid S) \sbis 
(P''_2 \mid S \mid S') \mid (Q \mid S \mid S')$.
To summarise, we have shown that 
$(P_2\mid Q \mid S) \wact{\tau} (P'''_2 \mid Q \mid S) \susp$,
\[
\begin{array}{l}
(P_1\mid Q \mid S) \sbis 
(P_1\mid S \mid S') \mid (Q \mid S \mid S') \rel{R}
(P''_2 \mid S \mid S') \mid (Q \mid S \mid S') \sbis
(P'''_2 \mid Q \mid S), \mand \\

\lf P_1\mid Q \mid S \rf \sbis 
\lf P_1\mid S \mid S' \mid Q \mid S \mid S' \rf \rel{R}
\lf P''_2 \mid S \mid S' \mid Q \mid S \mid S' \rf \sbis
\lf P'''_2 \mid Q \mid S \rf
\end{array}
\]
as required by the notion of labelled bisimulation up to $\sbis$.

\Proofitem{(L2)}
Suppose $P_1\mid Q \act{s} P'_1\mid Q$.

\Defitem{(L2)[1]} Suppose $P_1\act{s} P'_1$.

\Defitem{(L2)[1.1]}
If $P_2 \wact{s} P'_2$ and $P'_1\lbis P'_2$ we are done.

\Defitem{(L2)[1.2]}
If $P_1\wact{\tau} P'_2$ and $P'_1\lbis P'_2\mid (\s{emit} \ s)$
then $P_2\mid Q \wact{\tau} P'_2 \mid Q$ and 
we note that $(P'_2\mid Q) \mid (\s{emit} \ s)\sbis 
(P'_2 \mid (\s{emit} \ s))\mid Q$. 

\Defitem{(L2)[2]}
Suppose $Q\act{s} Q'$. Then $(P_2\mid Q) \act{s} (P_2\mid Q')$ and we
are done.

\Proofitemb{(5)}
Let $Q=\s{present} \ s\ P \ B$ and 
$Q'=\s{present} \ s\ P' \ B'$.

\Proofitem{(B1)} Note that $\neg (Q \act{\tau} \cdot)$.


\Proofitem{(B3)} Note that $\neg (Q \act{\ol{s}} \cdot)$.

\Proofitem{(L1)} Suppose $S=\s{emit} \ s_1 \mid \cdots \mid \s{emit} \
s_n$ and that $(Q\mid S) \susp$. Then $s_i\neq s$ for $i=1,\ldots,n$
and $\lf Q \mid S \rf = \lc B \rc_{\set{s_1,\ldots,s_n}}$. Note that
$(Q'\mid S) \susp$ too, and from the hypothesis
$B \lbis B'$ we derive
$\lf Q \mid S \rf \lbis \lf Q'\mid S \rf = 
\lc B \rc_{\set{s_1,\ldots,s_n}}$.

\Proofitem{(L2)}
The transition $\s{present} \ s \ P \ B \act{s} P\mid (\s{emit} \ s)$
is matched by $\s{present} \ s \ P' \ B' \act{s} P'\mid (\s{emit} \ s)$.
By hypothesis, $P\lbis P'$ and by (1), we derive 
$P\mid  (\s{emit} \ s) \lbis P'\mid  (\s{emit} \ s)$. \qed

\subsection{Proof lemma \ref{tau-lemma}}\label{proof-tau-lemma}
\Proofitemf{(1)}
Condition $(B3)^{\susp}$ is weaker 
than condition $(B3)$. Therefore, $P\lbis Q$ implies $P\lbissusp Q$.

\Proofitem{(2)} Reflexivity is obvious. For transitivity,
as usual, we have to check that $\lbissusp \comp \lbissusp$ is 
a $\susp$-labelled bisimulation. 
We focus on the new condition $(B3)^{\susp}$.
Suppose $P_1 \lbissusp P_2 \lbissusp P_3$, $P_1 \susp$, and
$P_1\act{\ol{s}}\cdot$. By $(B3)^{\susp}$, 
$P_2\wact{\tau} P'_2$ and $P'_2\act{\ol{s}}\cdot$.
By $(B2)$, $P_2\wact{\tau} P''_2$, $P''_2\susp$,
and $P_1\lbissusp P''_2$. By confluence,
$P'_2 \wact{\tau} P''_2$ and $P''_2 \act{\ol{s}}\cdot$.
By $(B1)$, $P_3 \wact{\tau} P'_3$ and 
$P''_2 \lbissusp P'_3$. By $(B3)^\susp$, $P'_3\wact{\tau}
P''_3$, $P''_2 \lbissusp P''_3$, and $P''_3 \act{\ol{s}}\cdot$.
Thus we have that $P_3\wact{\tau} P''_3$,  
$P''_3\act{\ol{s}}\cdot$, and
$P_1 \lbissusp P''_2 \lbissusp P''_3$ as required by
condition $(B3)^\susp$.

\Proofitem{(3)} We check that:
\[
\cl{R}=\w{Id} \union \set{(P,Q) \mid 
               P\act{\tau}Q \mbox{ or }
               Q \act{\tau} P}
\]
is a labelled bisimulation up to $\sbis$, where $\w{Id}$ is 
the identity relation. Thus $P\act{\tau} Q$ implies
$P\lbis Q$. By (1), $P\lbissusp Q$ and by proposition \ref{tr-lbis1},
$\w{tr}(P)=\w{tr}(Q)$.

\Defitem{(B1)} Suppose $P\act{\tau}P_1$. 
If $P\act{\tau} Q$ then by confluence, either 
$P_1=Q$ or 
$\xst{P_{12}}{P_1\act{\tau}P_{12} \mand Q\act{\tau} P_{12}}$.
In the first case, $Q\wact{\tau}Q$ and $(P_1,Q)\in\cl{R}$.
In the second case, $Q\wact{\tau}P_{12}$ and $(P_1,P_{12})\in \cl{R}$.
On the other hand, if $Q\act{\tau} P$ then $Q \wact{\tau} P_1$.


\Defitem{(B3)} Suppose $P\lsusp$ and $P\act{\ol{s}}\cdot$.
If $P\act{\tau}Q$ then $Q\act{\ol{s}}\cdot$ and $Q\wact{\tau} Q$.
On the other hand, if 
$Q\act{\tau} P$ then $Q\wact{\tau} P$.

\Defitem{(L1)}
If $P\act{\tau} Q$ then $P\mid S\susp$ is impossible.
On the other hand, if 
$Q\act{\tau} P$ and  $P\mid S \susp$ then 
$Q\mid S \wact{\tau} P\mid S$.

\Defitem{(L2)}
Suppose $P\act{s} P_1$.
If $P\act{\tau} Q$  then either $P_1=Q$ or 
$\xst{P_{12}}{P_1\act{\tau} P_{12} \mand Q \act{s} P_{12}}$.
In the first case, we have $Q\wact{\tau}Q$ and 
$P_1\  \cl{R}\  Q \sbis Q\mid (\s{emit} \ s)$.
In the second case, $Q\wact{s}P_{12}$ and $(P_1,P_{12})\in \cl{R}$.
On the other hand, if 
$Q\act{\tau} P$ then $Q\wact{s} P_1$.

\Proofitem{(4)}
Obviously, the critical condition to check is $(B3)$.
By proposition \ref{convergence-prop} we can use the predicate
$\wsusp$ rather than the predicate $\lsusp$.
So suppose $P_1\lbis Q_1$, $P_1\act{\ol{s}}\cdot$, $P_1\wact{\tau} P_2$, 
and $P_2\susp$. 
By $(B1)$, $Q_1\wact{\tau}Q_2$ and $P_2\lbissusp Q_2$.
By $(B3)^{\susp}$, 
$Q_2\wact{\tau} Q_3$, $Q_3\act{\ol{s}}\cdot$, and $P_2\lbissusp Q_3$.
By (3), $P_1 \lbissusp P_2$.
By transitivity of $\lbissusp$, $P_1 \lbissusp Q_3$. \qed

\subsection{Proof of lemma \ref{tr-lbis2}}\label{proof-tr-lbis2}
\Proofitemf{(1)} This follows from the remark that 
$P\mid (\s{emit} \ s) \act{I/O} P'$ if and only if 
$P\act{I\union\set{s}/O} P'$.

\Proofitem{(2)} We check the 5 conditions.

\Defitem{(B1)}
If $P\act{\tau} P'$ then $\w{tr}(P)=\w{tr}(P')$,
by lemma \ref{tau-lemma}(3). Thus
$Q\wact{\tau}Q$ and $(P',Q)\in\cl{R}$.


\Defitem{(B3)}
In view of proposition \ref{ddagger-bis}, it is enough to check condition
$(B3)^{\susp}$.
If $P\susp$ and $P\act{\ol{s}}\cdot$ then $P\act{\emptyset/O}\lf P
\rf$ and $s\in O$.
Thus $Q\act{\emptyset/O}Q'$. In particular, 
$Q\wact{\tau} Q''$, $Q''\act{\ol{s}}$.
By lemma \ref{tau-lemma}(3), $\w{tr}(Q)=\w{tr}(Q'')$.
Thus $(P,Q'')\in\cl{R}$.

\Defitem{(L1)}
If $P\mid S\susp$ then $P\act{I/O} P'$ 
where $I=\set{s \mid S\act{\ol{s}}\cdot}$, 
$P\mid S \wact{\tau} P''$, $P''\susp$, 
$O=\set{s \mid P''\act{\ol{s}}\cdot}$, and 
$P'=\lf P''\rf$.
By (1), $\w{tr}(P\mid S) = \w{tr}(Q\mid S)$.
Thus $Q\act{I/O} Q'$ where
$Q\mid S \wact{\tau} Q''$, $Q''\susp$,
and $Q'=\lf Q''\rf$.
Now $(P'',Q''), (P',Q')\in \cl{R}$ since
by lemma \ref{tau-lemma}(3)
$\w{tr}(P'')= \w{tr}(P\mid S) = \w{tr}(Q\mid S) = \w{tr}(Q'')$.

\Defitem{(L2)}
If $P\act{s} P'$ then 
$(P\mid (\s{emit} \ s)) \act{\tau} (P'\mid (\s{emit} \ s))$ and 
by lemma \ref{tau-lemma}(3)
$\w{tr}(P\mid \ol{s}) = \w{tr}(P' \mid (\s{emit} \ s))$.
Moreover, $P'\lbis (P'\mid (\s{emit} \ s))$ thus
by proposition \ref{tr-lbis1}, 
$\w{tr}(P')= \w{tr}(P'\mid (\s{emit} \ s))$.
By (1), $\w{tr}(P \mid (\s{emit} \ {s}))= 
\w{tr}(Q\mid (\s{emit} \ s))$.
We can conclude by considering  that
$Q\wact{\tau} Q$ and 
$(P',Q \mid (\s{emit} \ s))\in \cl{R}$ 
since $\w{tr}(P') = \w{tr}(P'\mid (\s{emit} \ s)) = 
\w{tr}(P\mid (\s{emit} \ s)) = \w{tr}(Q\mid (\s{emit} \ s))$. \qed

\end{document}